\documentclass{emulateapj}

\begin{document}

\shortauthors{Luhman et al.}
\shorttitle{Disk Population of Chamaeleon~I}

\title{The Disk Population of the Chamaeleon~I Star-Forming 
Region\altaffilmark{1}}

\author{
K. L. Luhman\altaffilmark{2},
L. E. Allen\altaffilmark{3},
P. R. Allen\altaffilmark{2},
R. A. Gutermuth\altaffilmark{3},
L. Hartmann\altaffilmark{4},
E. E. Mamajek\altaffilmark{3},
S. T. Megeath\altaffilmark{5},
P. C. Myers\altaffilmark{3},
and G. G. Fazio\altaffilmark{3}}

\altaffiltext{1}
{Based on observations performed with the Magellan Telescopes 
at Las Campanas Observatory, Gemini Observatory, and the {\it Spitzer Space
Telescope}. Gemini Observatory is operated by AURA under a
cooperative agreement with the NSF on behalf of the Gemini partnership: the
NSF (United States), the Particle Physics and Astronomy
Research Council (United Kingdom), the National Research Council (Canada),
CONICYT (Chile), the Australian Research Council (Australia), CNPq (Brazil)
and CONICET (Argentina). {\it Spitzer} is operated by the Jet Propulsion 
Laboratory (JPL), California Institute of Technology under NASA contract 1407.  
Support for this work was provided by NASA through contract 1256790 issued 
by JPL. Support for the IRAC instrument was provided by NASA through 
contract 960541 issued by JPL. 
}

\altaffiltext{2}{Department of Astronomy and Astrophysics, The Pennsylvania
State University, University Park, PA 16802; kluhman@astro.psu.edu.}

\altaffiltext{3}{Harvard-Smithsonian Center for Astrophysics, Cambridge, 
MA 02138.}

\altaffiltext{4}{Department of Astronomy, The University of Michigan,
Ann Arbor, MI 48109.}

\altaffiltext{5}{Department of Physics and Astronomy, The University of Toledo,
Toledo, OH 43606.}

\begin{abstract}

We present a census of circumstellar disks in the Chamaeleon~I star-forming 
region. Using the Infrared Array Camera and the Multiband Imaging Photometer
onboard the {\it Spitzer Space Telescope}, we have obtained images of 
Chamaeleon~I at 3.6, 4.5, 5.8, 8.0, and 24~\micron.
To search for new disk-bearing members of the cluster, we have performed 
spectroscopy on objects that have red colors in these data.
Through this work, we have discovered four new members of Chamaeleon~I with 
spectral types of M4, M6, M7.5, and L0. 
The first three objects are highly embedded ($A_J\sim5$) and reside
near known protostars, indicating that they may be among the youngest 
low-mass sources in the cluster ($\tau<1$~Myr). 
The L0 source is the coolest known member of Chamaeleon~I. 
Its luminosity implies a mass of 0.004-0.01~$M_\odot$, making it the least 
massive brown dwarf for which a circumstellar disk has been reliably detected. 
To characterize the disk population in Chamaeleon~I, we have 
classified the infrared spectral energy distributions of the 203 known members 
that are encompassed by the {\it Spitzer} images.
Through these classifications, we find that the disk fraction in Chamaeleon~I
is roughly constant at $\sim50$\% from 0.01 to 0.3~$M_\odot$. These data
are similar to the disk fraction of IC~348, which is a denser cluster
at the same age as Chamaeleon~I.
However, the disk fraction at $M\gtrsim1$~$M_\odot$ is significantly higher 
in Chamaeleon~I than in IC~348 (65\% vs.\ 20\%), indicating
longer disk lifetimes in Chamaeleon~I for this mass range.
Thus, low-density star-forming regions like Chamaeleon~I may offer more time
for planet formation around solar-type stars than denser clusters.

\end{abstract}


\section{Introduction}
\label{sec:intro}

Low-mass stars and brown dwarfs are important sites for studies of
planet formation because most stars in the galaxy are low-mass stars 
while brown dwarfs offer an opportunity to explore planet formation in
an extreme environment. 
Planet formation around low-mass stars and brown dwarfs can be investigated 
through observations of their primordial circumstellar disks.
These observations are most easily performed at mid-infrared 
(IR) wavelengths ($\lambda\sim5$-20~\micron) because this wavelength range 
provides the best combination of contrast of the disk relative to 
stellar photosphere, disk brightness, and sensitivity of available telescopes.

Molecular clouds that contain ongoing star formation ($\tau\lesssim5$~Myr) 
are the natural laboratories for observations of circumstellar disks. 
Among these regions, Chamaeleon~I is arguably the best site for studying 
disks around low-mass stars and brown dwarfs. It is one of the nearest 
star-forming regions to the Sun \citep[$d=160$-170~pc,][]{whi97,wic98,ber99}, 
making its low-mass members relatively bright. The cluster is young enough 
that it retains a significant population of primordial disks, but it is old
enough that most of its members are no longer highly obscured by dust 
($A_V\lesssim5$). Because of the relatively low extinction, 
optical wavelengths are accessible for the spectral classification of 
the stellar population \citep[][references therein]{com04,luh04cha} 
and for measuring diagnostics of accretion \citep{moh05,muz05}. 
Optical and near-IR imaging and spectroscopic surveys of Chamaeleon~I
have produced an extensive census of the stellar and substellar cluster
members \citep{luh07cha}. In particular, this census is unbiased in terms of 
disks, which is essential for measuring the prevalence of disks. 
With $\sim200$ known members at masses of 0.01-1~$M_\odot$, 
Chamaeleon~I is sufficiently rich for a statistical analysis of disks around
low-mass stars and brown dwarfs.  
This region is also well-suited for observations of disks at mid-IR 
wavelengths. The stellar population is sparse enough that current 
mid-IR telescopes can resolve individual members, while compact enough for
deep imaging of a large fraction of the cluster in a reasonable amount
of time. In addition, the cloud exhibits relatively little nebulosity 
and extended emission at mid-IR wavelengths, which enables accurate 
photometry of the faint, low-mass members.

Several studies have searched for evidence of disks in Chamaeleon~I 
through near- and mid-IR photometry. 
During pointed observations of known members, \citet{glass79} and 
\citet{jay03} obtained photometry in the $JHKL$ bands
(1.2, 1.6, 2.2, 3.4~\micron) for 17 and 15 members with spectral
types as late as M4 and M8, respectively. Using a similar set of filters,
\citet{gk01} and \citet{kg01} performed wide-field imaging of a 0.5~deg$^2$
field that encompassed a large fraction of the Chamaeleon~I cloud.
Photometry at wavelengths longward of the $L$-band offers more reliable 
detections of disks because of the better contrast of 
cool circumstellar dust relative to warmer stellar photopsheres. 
To obtain data of this kind for known young stars in Chamaeleon~I and 
to search for new disk-bearing members of the cluster, 
\citet{bau84} observed Chamaeleon~I with
the {\it Infrared Astronomical Satellite} ({\it IRAS}) at 12, 25, 60, and 
100~\micron. They detected 70 sources, half of which are cluster members.
A mid-IR survey with better sensitivity and spatial resolution 
was later conducted with the {\it Infrared Space Observatory} 
\citep[{\it ISO},][]{nor96,per00}, which provided photometry for 99 known
members, including a brown dwarf with a spectral
type near M8 \citep{nt01,apa02}. Mid-IR imaging also has been performed
toward smaller areas surrounding the three most prominent reflection nebulae
in the Chamaeleon~I cloud
using {\it IRAS} \citep{ass90,pru91} and {\it ISO} \citep{leh01}.

The {\it Spitzer Space Telescope} \citep{wer04} is currently the most
advanced telescope operating at mid-IR wavelengths. 
Because of its high spatial resolution, excellent sensitivity, and large 
field of view, {\it Spitzer} is proficient at identifying 
disk-bearing members of young clusters \citep{all04,gut04,meg04,muz04}. 
This capability has been widely utilized through {\it Spitzer} surveys
of nearby star-forming regions and young associations ($\tau\lesssim10$~Myr),
including Taurus \citep{har05,luh06tau2,gui07}, 
Perseus \citep{luh05frac,lada06,jor06,mue07,reb07,gut07}, 
Lupus \citep{allen07}, Serpens \citep{har06,har07,win07}, 
$\sigma$~Ori \citep{her07,cab07,zap07,sch07b}, $\lambda$~Ori \citep{bar07},
Orion OB1b \citep{her06}, $\eta$~Cha \citep{meg05}, 
Chamaeleon~II \citep{you05,all06,por07}, NGC~2362 \citep{dahm07}, 
Upper Sco \citep{car06,sch07a}, Tr 37, and NGC~7160 \citep{sic06}.
Extensive {\it Spitzer} imaging has been performed in Chamaeleon~I as well.
Some of these data have been used to examine the properties of the
disk population of Chamaeleon~I \citep{luh05frac,dam07}, identify 
low-mass brown dwarfs with disks \citep{luh05ots,luh05cha}, and 
search for outflows \citep{bal06}. 
In this paper, we present a comprehensive census of disks in Chamaeleon~I
using most of the {\it Spitzer} images between 3.6 and 24~\micron\ that have 
been obtained in this region. 
We begin by describing the collection and analysis of the 
{\it Spitzer} images (\S~\ref{sec:images}) and the identification of 
new disk-bearing members of the cluster with these data (\S~\ref{sec:new}).
We then use the {\it Spitzer} photometry of the known members of the cluster
to investigate the global properties of the disk population of Chamaeleon~I 
(\S~\ref{sec:global}).

\section{Infrared Images}
\label{sec:images}

\subsection{Observations}

For our census of the disk population in Chamaeleon~I, we use 
images at 3.6, 4.5, 5.8, and 8.0~\micron\ obtained with 
{\it Spitzer}'s Infrared Array Camera \citep[IRAC;][]{faz04a} and 
images at 24~\micron\ obtained with the Multiband Imaging Photometer for
{\it Spitzer} \citep[MIPS;][]{rie04}.
The fields of view are $5.2\arcmin\times5.2\arcmin$ and 
$5.4\arcmin\times5.4\arcmin$ for IRAC and the 24~\micron\ channel of MIPS, 
respectively. The cameras produce images with FWHM=1.6-$1.9\arcsec$ from 3.6 
to 8.0~\micron\ and FWHM=$5.9\arcsec$ at 24~\micron.

\begin{figure}
\epsscale{1}
\plotone{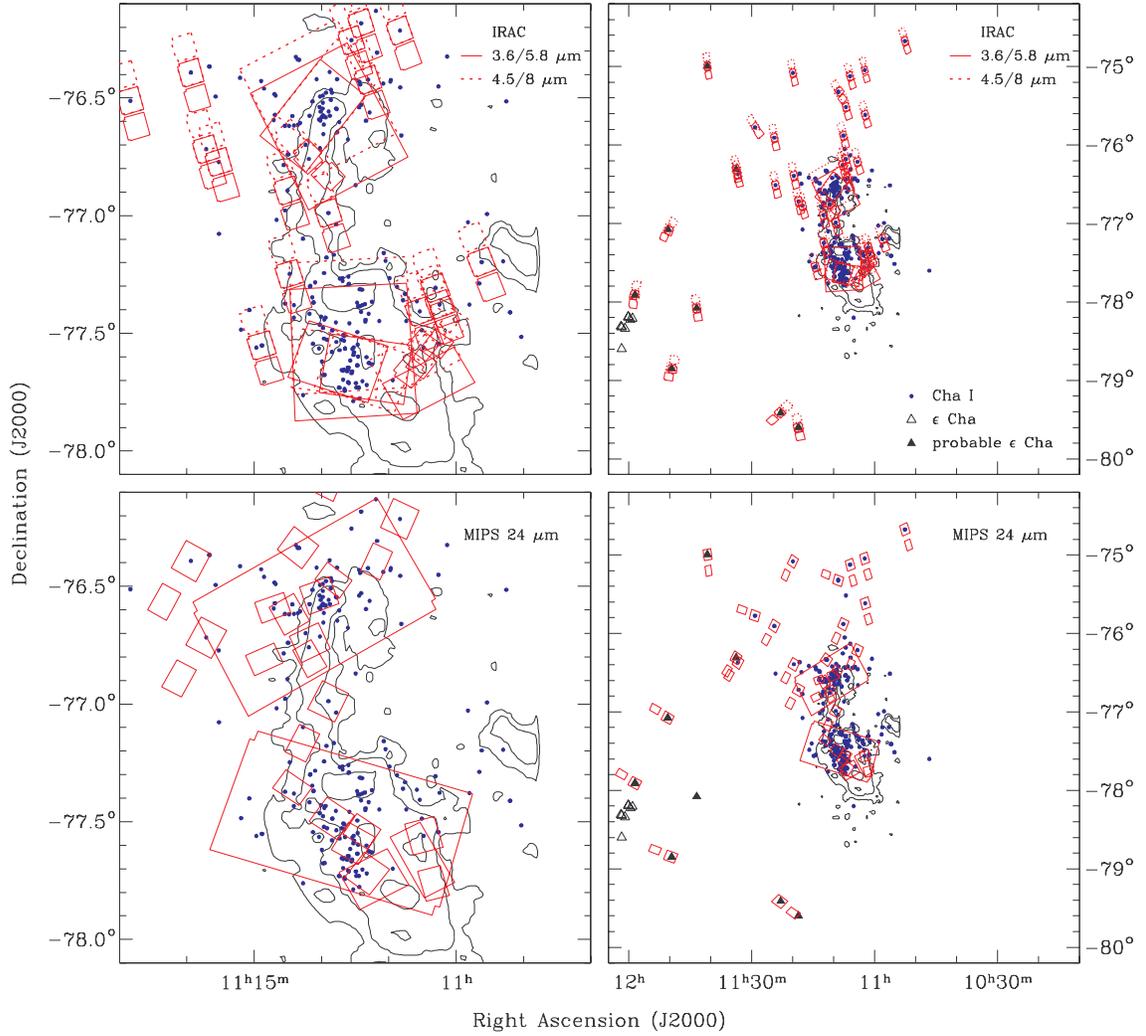}
\caption{
Fields in the Chamaeleon~I star-forming region that have been imaged
with IRAC ({\it top}) and MIPS ({\it bottom}). The known members of the 
cluster are indicated ({\it points}). We also have measured 
IRAC and MIPS photometry for young stars near Chamaeleon~I that we have 
identified as probable members of the $\epsilon$~Cha young association 
(\S~\ref{sec:app}, {\it filled triangles}). 
The original members of $\epsilon$~Cha are located $3\arcdeg$ southeast
of the Chamaeleon~I clouds \citep[{\it open triangles},][]{fei03,luh04eta}.
The contours represent the extinction map of \citet{cam97} at intervals
of $A_J=0.5$, 1, and 2.
}
\label{fig:map}
\end{figure}

In this paper, we consider all IRAC and MIPS 24~\micron\ observations
that have been performed within a radius of $3\arcdeg$ from 
$\alpha=11^{\rm h}07^{\rm m}00^{\rm s}$, $\delta=-77\arcdeg10\arcmin00\arcsec$
(J2000) with the exception of the recently executed Legacy program of L. Allen,
which has a program identification (PID) of 30574. These data were obtained 
through IRAC Guaranteed Time Observations (GTO) for PID=6, 36, and 37 
(G. Fazio), IRAC and IRS GTO for PID=30540 (G. Fazio, J. Houck), IRS GTO for 
PID=40302 (J. Houck), Director's Discretionary Time for PID=248 (S. Mohanty), 
and Legacy programs for PID=139, 173 (N. Evans), and 148 (M. Meyer). 
The characteristics of the IRAC and MIPS images are summarized in 
Tables~\ref{tab:iraclog} and \ref{tab:mipslog}, respectively. 
The boundaries of these images are indicated in the maps
of Chamaeleon~I in Figure~\ref{fig:map}.

\subsection{Data Reduction}
\label{sec:reduction}

Initial processing of the IRAC images was performed with the Spitzer Science 
Center (SSC) S14.0.0 pipeline. The resulting Basic Calibrated Data (BCD) 
images were automatically treated for bright source artifacts 
(jailbar, pulldown, muxbleed, and  banding) with
customized versions of IDL routines developed by the IRAC instrument team
\citep{hor04,pip04}. Mosaicking of the treated BCDs was
performed using R. Gutermuth's WCSmosaic IDL package, which includes the
following features: a redundancy-based cosmic ray detection and rejection
algorithm; frame-by-frame distortion correction, derotation, and subpixel
offsetting in a single transformation (to minimize data smoothing); and
on-the-fly background matching. WCSmosaic was built with heavy dependence on
the FITS and WCS access and manipulation procedures provided by the IDL
Astronomy Users Library \citep{lan93}.
We selected a plate scale of $0.86\arcsec$~pixel$^{-1}$ for the reduced IRAC 
mosaics, which is the native scale divided by $\sqrt{2}$.

For the processing of the MIPS data, the SSC pipeline versions were S16.0.1 
for Astronomical Observation Request (AOR) identifications of 5687040, 
5688064, 5688320, and 5688576 and S16.1.0 for all other AORs.
The resulting 24~\micron\ images were reduced further 
with the MOPEX software from the SSC. The plate scale
of the final images was $2.45\arcsec$~pixel$^{-1}$.

We used the IRAF task STARFIND to identify all point sources in the reduced 
IRAC and MIPS images. From the initial list of 3.6~\micron\ detections from 
STARFIND, we rejected sources that were not detected in the 4.5~\micron\ image
for areas that were imaged by both channels. Similarly, we rejected
STARFIND detections at 4.5~\micron\ that did not have a corresponding detection
at 3.6~\micron. For the data at 5.8 and 8.0~\micron, we retained only the 
sources that were detected in the 3.6 and 4.5~\micron\ images, respectively. 
We marked the sources produced by these criteria on the images and removed 
remaining spurious detections through visual inspection.

We measured aperture photometry for the IRAC and MIPS sources using the 
IRAF task PHOT. We adopted zero point magnitudes ($ZP$) of 19.670, 18.921, 
16.855, 17.394, and 15.119 in the 3.6, 4.5, 5.8, 8.0, and 24~\micron\ bands, 
where $M=-2.5 \log (DN/sec) + ZP$ \citep{rea05,eng07}.  
For IRAC, we used an aperture radius of 4 pixels and inner and outer 
radii of 4 and 5 pixels, respectively, for the sky annulus.
For MIPS, we adopted radii of 3, 3, and 4 pixels for the aperture and
the inner and outer boundaries of the sky annulus, respectively.
Using bright, isolated stars in the IRAC and MIPS images, we measured
aperture corrections between our adopted apertures and
the larger ones employed by \citet{rea05} and \citet{eng07}. For the MIPS
data, we combined our aperture correction with the one estimated by 
\citet{eng07} between their aperture and an infinite one. 
The total aperture corrections applied to our measurements are 0.18, 0.17, 
0.23, 0.48, and 0.8~mag for 3.6, 4.5, 5.8, 8.0, and 24~$\mu$m, respectively. 
Smaller apertures (and the appropriate aperture corrections) were used for 
sources that were near other stars, including the known members Hn~21W 
and E, the new member 2MASS~J11062942$-$7724586 and its candidate companion 
Cha~J11062788$-$7724543, and the new member 2MASS~J11084296$-$7743500, which is
close to the protostar IRN. Our quoted photometric errors 
include the Poisson errors in the source and background emission 
and the 2\% and 4\% uncertainties in the calibrations of IRAC and MIPS,
respectively \citep{rea05,eng07}. 
The errors do not include an additional error of $\pm0.05$~mag due to 
location-dependent variations in the IRAC calibration.

\begin{figure}
\epsscale{1}
\plotone{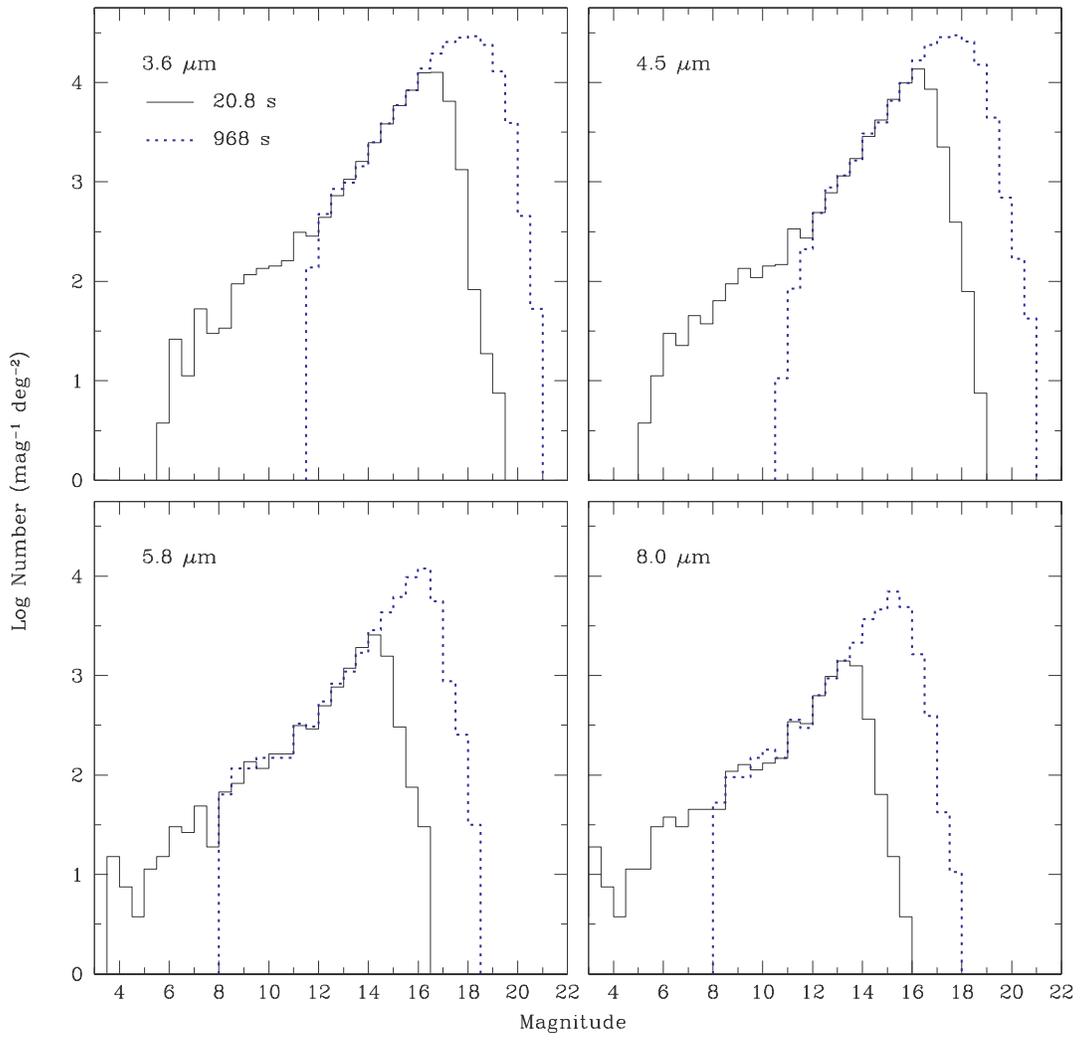}
\caption{
Distributions of IRAC magnitudes for images of Chamaeleon~I that have
short ({\it solid lines}) and long ({\it dotted lines}) total exposure times.
}
\label{fig:limit}
\end{figure}

To illustrate the detection and completeness limits of the IRAC data,
we consider the two shallow maps (AOR=3960320, 3651328) and two of the three
deep maps (AOR=3955968, 12620032) that are centered on the northern and
southern subclusters. The total exposure times at a given position in
these maps are 20.8 and 968~s, respectively. For each exposure time and filter, 
we estimate the completeness limit by measuring the magnitude
at which the logarithm of the number of sources as a function of magnitude
departs from a linear slope and begins to turn over.
Based on the distributions of IRAC magnitudes, which are shown in 
Figure~\ref{fig:limit}, we arrive at completeness limits of 
16.25, 16.25, 14.25, and 13.5 for the short exposures and 
17.0, 16.75, 16.0, and 15.25 for the long exposures at 3.6, 4.5, 5.8, and 
8.0~\micron, respectively.
The estimates for the short exposures are confirmed by the comparisons of
the histograms for the short and long exposures in Figure~\ref{fig:limit}.
The improvement in the completeness limits from the short exposures to
the long ones is smaller at 3.6 and 4.5~\micron\ than at 5.8 and 
8.0~\micron\ because the short exposures at 3.6 and 
4.5~\micron\ are already near the depth at which source confusion 
begins to reduce completeness \citep{faz04b}. 
Thus, the completeness limits of the longer exposures at 3.6 and 
4.5~\micron\ do not scale with $\sqrt{\tau}$ and instead are limited by 
confusion. Our limits at 3.6 and 4.5~\micron\ are similar to those measured by
\citet{faz04b} for deeper images of extragalactic fields.
The short exposures at 8.0~\micron\ exhibit the brightest completeness 
limit in our data, corresponding to a mass of $\sim$0.015~$M_\odot$ 
and a spectral type of $\sim$M9 for an age of 2~Myr and the distance of
Chamaeleon~I according to evolutionary models \citep{cha00}.
The fainter completeness limits of the other filters and exposure times are
equivalent to masses of $\sim$0.003-0.01~$M_\odot$ for members of Chamaeleon~I.

\section{New Members of Chamaeleon I}
\label{sec:new}

\subsection{Selection of Candidate Members}
\label{sec:select}

\begin{figure}
\epsscale{0.6}
\plotone{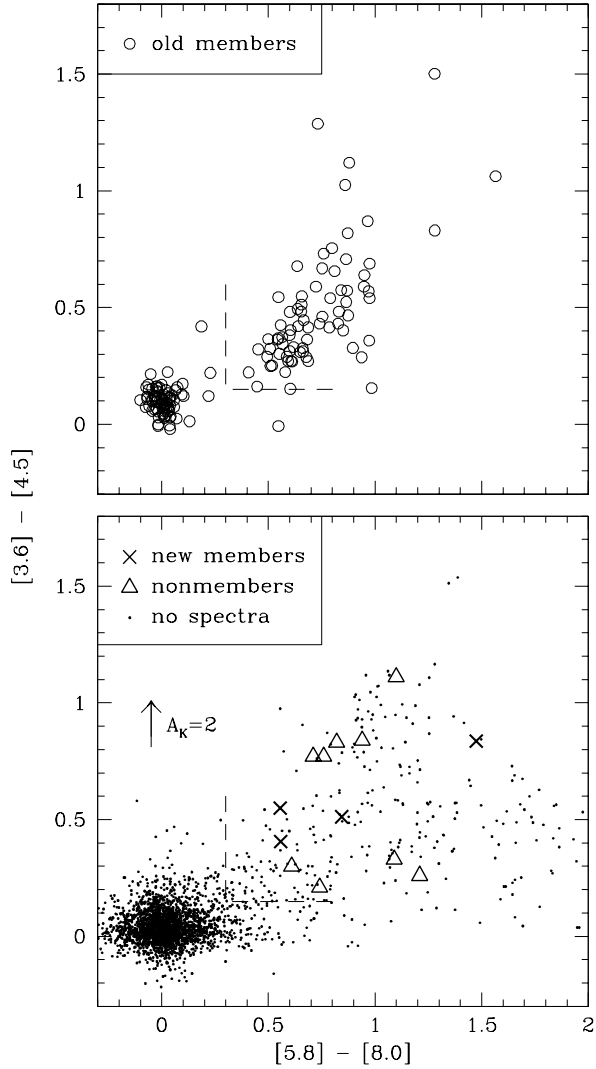}
\caption{
{\it Spitzer} IRAC color-color diagrams for the survey fields in the
Chamaeleon~I star-forming region that are indicated in Figure~\ref{fig:map}.
{\it Top}: The previously known members of Chamaeleon~I ({\it circles})
exhibit either neutral colors consistent with stellar photospheres or
significantly redder colors that are indicative of circumstellar disks.
{\it Bottom}: Among the other point sources in the IRAC data,
we selected a sample of candidate disk-bearing members based on colors
of $[5.8]-[8.0]>0.3$ and $[3.6]-[4.5]>0.15$ ({\it dashed line}) and
obtained spectra of them. The candidates classified as
new members ({\it crosses}) and nonmembers ({\it triangles}) are listed in
Tables~\ref{tab:new} and \ref{tab:non}, respectively.
Most of the remaining red sources are probably galaxies based on their faint
magnitudes (see Fig.~\ref{fig:14}).
Only objects with photometric errors less than 0.1~mag in all four bands
are shown in these diagrams, with the exception of the new member 
2M~J1108$-$7743.
The reddening vector is based on the extinction law from \citet{fla07}.
}
\label{fig:1234}
\end{figure}

As shown in the maps of Chamaeleon~I in Figure~\ref{fig:map}, most
of the star-forming region has been imaged by IRAC and MIPS. 
Thus, we can use these data to search for new disk-bearing members of the 
cluster through the presence of mid-IR emission in excess above that expected 
from stellar photospheres. A color-color diagram constructed from 
IRAC measurements is an efficient tool for identifying sources 
of this kind \citep{all04,meg04}. 
Because our images encompass a large fraction of the known stellar
population of Chamaeleon~I, we can use our photometry for these sources to
delineate the typical IRAC colors of young stars with disks.
In Figure~\ref{fig:1234}, we plot a diagram of $[3.6]-[4.5]$ versus
$[5.8]-[8.0]$ for all known members that are not saturated and that have
photometric uncertainties less than 0.1~mag in all of the IRAC bands,
which consists of 153 sources.
If an object has multiple measurements available in a given band, we 
have adopted the average of those measurements. 
As in other star-forming regions \citep{har05}, the known members of 
Chamaeleon~I form two distinct populations in Figure~\ref{fig:1234},
stars clustering tightly near the origin and stars with much redder colors.
The former are consistent with stellar photospheres while the latter
indicate the presence of hot, optically-thick dust.
In surveys for new members of Taurus and Lupus, \citet{luh06tau2} and
\citet{allen07} used color criteria of $[3.6]-[4.5]>0.15$ and 
$[5.8]-[8.0]>0.3$ for selecting stars that might have disks.
These colors bound the disk population in Chamaeleon~I as well.
Thus, we have applied these criteria to all sources in our photometric 
catalog for Chamaeleon~I that have errors less than 0.1~mag in all four 
bands, which are shown in Figure~\ref{fig:1234}. Stars classified as 
nonmembers in previous studies have been omitted \citep{luh04cha,luh07cha}. 
This process produced 350 candidate disk-bearing sources. 
Although its photometric errors are larger than 0.1~mag, we have also 
identified 2MASS~J11084296$-$7743500 as a promising candidate because of
its red IRAC colors and its close proximity to the protostar IRN.

We selected 17 candidates for spectroscopy to assess their membership,
focusing on sources that are within the two main subclusters of Chamaeleon~I 
and that are bright enough for optical or near-IR spectroscopy. 
The spectroscopic observations of eight of these candidates were described
by \citet{luh07cha} while the data for the remaining nine candidates 
are presented in \S~\ref{sec:spectra}. Using these spectra, we classify 
four candidates as new members and 13 candidates as nonmembers.
One of the nonmembers, 2MASS~J11081204$-$7622124, falls outside of the 
boundaries of the color-color diagram in Figure~\ref{fig:1234}.
Three of the nonmembers, 2MASS J11085651$-$7645154, 2MASS J11105185$-$7642259,
and 2MASS J11105353$-$7726389, were identified as candidates because they 
exhibited 
excesses in preliminary photometric measurements that were used for selecting 
candidates for spectroscopy. However, upon further data analysis, we found
that those apparent excesses were caused by contamination from cosmic rays, 
and that the sources do not exhibit excesses in our final photometry.
For this reason, these three objects are excluded from Figure~\ref{fig:1234}.

\subsection{Spectroscopy of Candidates}
\label{sec:spectra}

\subsubsection{Observations}

We performed optical and near-IR spectroscopy on nine of the candidate 
members of Chamaeleon~I that were selected in \S~\ref{sec:select}
using the Low Dispersion Survey Spectrograph (LDSS-3) on the Magellan~II
Telescope and the Gemini Near-Infrared Spectrograph (GNIRS) at
Gemini South Observatory. The GNIRS observations were conducted through
program GS-2005A-C-13. Table~\ref{tab:speclog} summarizes the observing 
runs and instrument configurations for the spectra. In Table~\ref{tab:new},
we indicate the night on which each new member was observed. 
We also observed known late-type members of Chamaeleon~I for comparison to the 
candidates, obtaining LDSS-3 spectra of OTS~44 and Cha~J11091363$-$7734446 
and GNIRS spectra of T50, Hn~12W, Cha~H$\alpha$~11, and CHSM~17173. 
The procedures for the collection and reduction of the 
LDSS-3 and GNIRS spectra were similar to those described by
\citet{luh04tau} and \citet{luh04ots}, respectively. 

\subsubsection{Nonmembers}

Our spectroscopic targets are fainter than most of the known members of 
Chamaeleon~I in the IRAC bands, which indicates that they should be 
low-mass stars or brown dwarfs if they are members of the cluster.
Among the 17 candidates observed spectroscopically by \citet{luh07cha} 
and in this work, 13 sources do not exhibit the molecular absorption bands 
that characterize cool, low-mass objects. They also lack the hydrogen 
emission lines that are often seen in young stars with disks. 
Therefore, we classify these candidates as nonmembers. 
The spectral classifications of eight of these nonmembers were presented by
\citet{luh04cha}. The remaining five sources
(Cha J11081724$-$7631107, Cha J11083029$-$7733067, Cha J11084006$-$7627085, 
Cha J11090876$-$7626234, Cha J11120546$-$7631376)
are classified as $<$M0 based on the absence of molecular absorption bands. 
  
\subsubsection{New Member in Optical Sample}

\begin{figure}
\epsscale{0.6}
\plotone{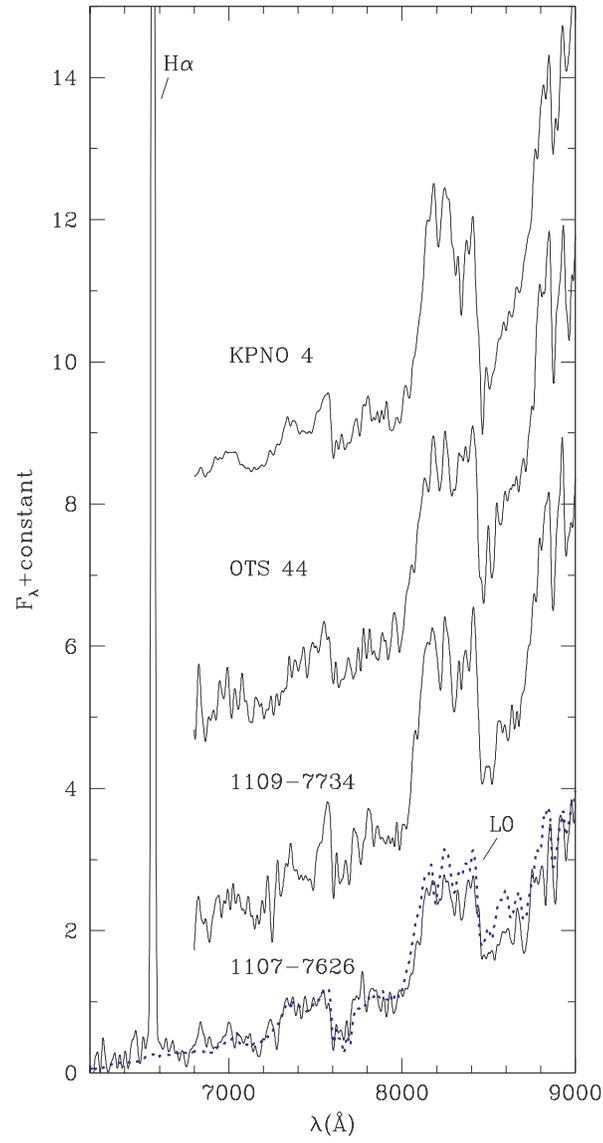}
\caption{
Optical spectrum of a new member of Chamaeleon~I, Cha~J1107$-$7626,
compared to data for the young M9.5 objects KPNO~4, OTS~44, and 
Cha~J11091363$-$7734446 \citep{bri02,luh07cha} and the L0 field 
dwarf 2MASS~J03454316+2540233 \citep[{\it dotted line},][]{kir99}.
The spectrum of Cha~J11091363$-$7734446 has been dereddened by $A_J=0.28$
\citep{luh07cha}.
The data are displayed at a resolution of 18~\AA\ and are normalized at
7500~\AA.
}
\label{fig:op}
\end{figure}

The remaining four candidates do show late-type spectral features and 
evidence of membership in Chamaeleon~I. 
One of these candidates is in the LDSS-3 sample and the other three objects 
are in the GNIRS sample. We first discuss the source observed by LDSS-3,
Cha~J11070768$-$7626326 (hereafter Cha~1107$-$7626).
The optical spectrum of this candidate is shown in Figure~\ref{fig:op}. 
The most obvious feature of this spectrum is the
strong H$\alpha$ emission line. The equivalent width of the line
is uncertain because the surrounding continuum is weak and poorly
measured, but it appears to have a value of several hundred angstroms.
The line is clearly stronger than the emission observed in active late-type 
field dwarfs \citep{giz02} and indicates the presence of accretion. 
The H$\alpha$ emission and the mid-IR excess emission together 
demonstrate the youth of Cha~1107$-$7626, and thus we conclude that 
it is a member of Chamaeleon~I. 

To measure its spectral type, we have compared Cha~1107$-$7626 
to optically-classified late-type members of Chamaeleon~I
and other star-forming regions \citep{luh04cha,luh07cha} as well as standard
field dwarfs \citep{kir99}. The results of our classification are illustrated
in Figure~\ref{fig:op}, where we compare Cha~1107$-$7626 to the young 
late-type objects KPNO~4, OTS~44, and Cha~J11091363$-$7734446 and the L0 field 
dwarf 2MASS~J03454316+2540233 \citep{kir99}. KPNO~4 and OTS~44 have optical
spectral types of M9.5 \citep{bri02,luh07cha} and Cha~J11091363$-$7734446
was classified as $\geq$M9 with IR spectra \citep{luh07cha}. 
We now classify the latter as M9.5$\pm$0.5 based on the similarity to KPNO~4 
and OTS~44 in the optical data in Figure~\ref{fig:op}.
Compared to the three young M9.5 objects, Cha~1107$-$7626 exhibits
weaker TiO and VO absorption, which suggests a later spectral type. 
Indeed, its spectrum agrees better with the L0 field dwarf, as shown in
Figure~\ref{fig:op}. Because the transition from M to L types is defined
by the disappearance of TiO absorption at 7000-7200~\AA\ \citep{kir99},
we need a spectrum with a higher signal-to-noise ratio at those wavelengths
to definitively classify Cha~1107$-$7626 as L-type. However, given
that its spectrum agrees significantly better with L0 than M9.5, we 
tentatively classify it as L0. 

The comparison of Cha~1107$-$7626 to the field dwarf in Figure~\ref{fig:op} 
indicates that it has negligible extinction ($A_J\lesssim0.3$). 
By combining its $J$-band magnitude with a bolometric correction for L0
\citep{dah02} and a distance modulus of 6.05 \citep{luh08cha}, we 
estimate a bolometric luminosity of $3.3\times10^{-4}$~$L_\odot$.
At this luminosity, Cha~1107$-$7626 is tied with Cha~J11083040$-$7731387 
\citep{luh07cha} as the least luminous known member of Chamaeleon~I.

\begin{figure}
\epsscale{1}
\plotone{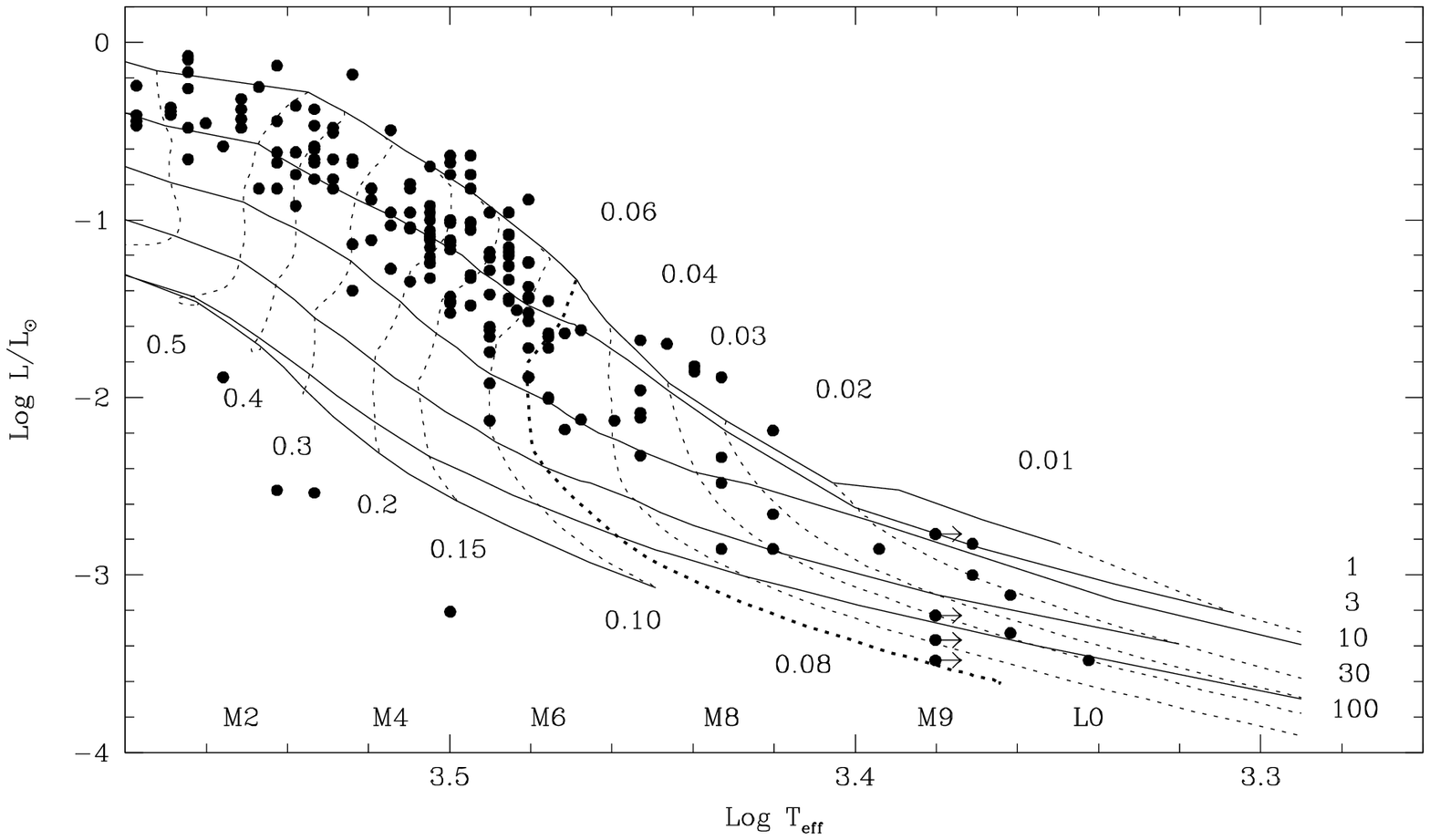}
\caption{
H-R diagram for all known low-mass stars and brown dwarfs in Chamaeleon~I
\citep[][\S~\ref{sec:new}]{luh07cha}.
These data are shown with the theoretical evolutionary models of
\citet{bar98} ($0.1<M/M_\odot\leq1$) and \citet{cha00} ($M/M_\odot\leq0.1$),
where the mass tracks ({\it dotted lines}) and isochrones ({\it solid lines})
are labeled in units of $M_\odot$ and Myr, respectively.
The four stars that are below the main sequence (ISO~225, CHSM~15991,
Cha~J11081938$-$7731522, ESO~H$\alpha$~569) are probably detected primarily 
in scattered light, which precludes the measurement of accurate luminosities.
}
\label{fig:hr}
\end{figure}

To estimate the mass of Cha~1107$-$7626, we begin by examining its position
on the Hertzsprung-Russell (H-R) diagram. 
To do this, we must convert the spectral type of Cha~1107$-$7626 to an 
effective temperature. 
In our previous studies of young late-type objects, we have performed 
conversions of this kind with the temperature scales from
\citet{luh99} and \citet{luh03ic}, which spanned M1 to M9.
The former scale was designed to produce coevality for the components of 
the young quadruple system GG~Tau (K7-M7.5) on the model isochrones
from \citet{bar98}.
\citet{luh03ic} adjusted that scale at M8 and M9 to improve the coevality
between the coolest members of Taurus and IC~348 and the members at earlier 
types. The resulting scale was then extrapolated to M9.25 and M9.5 
as cooler objects were discovered in Taurus and Chamaeleon~I
\citep{bri02,luh04ots,luh07cha}. For the purposes of this study, 
we continue this extrapolation to L0 for Cha~1107$-$7626, 
arriving at a value of 2200~K. 
Using this temperature and our luminosity estimate, we have
placed Cha~1107$-$7626 on the H-R diagram in Figure~\ref{fig:hr}. 
We have included all other known low-mass members of Chamaeleon~I
\citep{luh07cha} as well as the three new members that we have found with 
GNIRS (\S~\ref{sec:gnirs}). 
Four of the previously known members have formal spectral types of $\geq$M9
because they were classified through H$_2$O absorption bands, and the 
variation of these bands with optical spectral type is unknown for young 
objects later than M9.5. However, the strengths of the H$_2$O bands in these 
objects do agree with those of OTS~44 and KPNO~4, which have optical types
of M9.5. Thus, it is likely that they have spectral types near M9.5.

The cluster sequence for Chamaeleon~I in Figure~\ref{fig:hr} is parallel to 
the model isochrones for members earlier than M8, which is expected since the
adopted temperature scale was designed to produce coevality for the GG~Tau 
system. However, the sequence does not remain parallel to the isochrones 
at later types. Most members earlier than M8 appear between the isochrones for
1 and 10~Myr, while the coolest brown dwarfs, including 
Cha~1107$-$7626, are bracketed by 10 and 100~Myr. 
It is unlikely that these old ages are valid.
For instance, the lifetimes of molecular clouds are too short 
\citep[$\tau<10$~Myr,][]{har01} for the Chamaeleon~I clouds to have undergone
star formation across such an extended period of time. 
In addition, objects with ages of 10-100~Myr would have dispersed from the
cloud long ago. Instead, the old isochronal ages are probably a reflection of
errors in the adopted temperature scale and/or the evolutionary models.
If the models are valid, then the temperature scale for young objects at
M9-L0 must be much cooler than the adopted one, which is already cooler than 
the scale measured for field dwarfs \citep{kir05}.
Significant errors are probably present in both the temperature scale, 
which is unconstrained at the latest types, and the predicted effective 
temperatures, which are sensitive to various details of the models, such 
as the treatment of convection \citep{bar02}.
Compared to effective temperature, bolometric luminosity appears to be 
less sensitive to model uncertainties, at least for ages of $\tau\gtrsim1$~Myr
\citep{bar02}, and is easier to measure. 
Therefore, we have chosen to estimate the mass of Cha~1107$-$7626 from its 
luminosity. If we assume that it has an age in the range of values exhibited
by the stellar members of Chamaeleon~I \citep[$\tau=1$-6~Myr,][]{luh07cha}, 
then we arrive at a mass of 0.004-0.01~$M_\odot$ for Cha~1107$-$7626 
based on the luminosities predicted by \citet{bur97} and \citet{cha00}.
\citet{zap07} and \citet{sch07b} have reported detections of disks for
brown dwarfs at comparable masses in the $\sigma$~Ori cluster using IRAC data. 
However, because of the very large uncertainties in the 5.8 and 
8.0~\micron\ photometry in those studies (0.3-0.5~mag), their detections of
excess emission have low significance. 
Thus, Cha~1107$-$7626 is the least massive brown dwarf that has definitive
evidence for a circumstellar disk.

\subsubsection{New Members in Infrared Sample}
\label{sec:gnirs}

In the GNIRS sample, the three candidates that show evidence of membership
in Chamaeleon~I are 2MASS J11084296$-$7743500, 2MASS J11062942$-$7724586, and
2MASS J11070369$-$7724307 (hereafter 2M~1108$-$7743, 2M~1106$-$7724, 
2M~1107$-$7724).
The near-IR spectra of these objects exhibit strong H$_2$O absorption bands,
which indicates that they are low-mass stars or brown dwarfs rather 
than galaxies. High levels of reddening in their spectra demonstrate that
they are not field stars in the foreground of the cluster and 
they are too bright to be background field dwarfs. We also detect emission
in H$_{2}$~1-0~S(1) at 2.12~\micron\ in 2M~1108$-$7743, which is a signature
of very young stars. 
Finally, the spectral features that are sensitive to surface gravity in
our data, such as the shape of the $H$- and $K$-band continua 
\citep{luc01,luh04ots,kir06}, are inconsistent with field dwarfs and giants. 
Instead, the spectra of 2M~1108$-$7743, 2M~1106$-$7724, and 2M~1107$-$7724 
agree well with known cluster members observed with GNIRS in terms of the 
gravity-sensitive features. Based on these properties and the presence of 
mid-IR excess emission, we classify these sources as members of Chamaeleon~I.

\begin{figure}
\epsscale{0.6}
\plotone{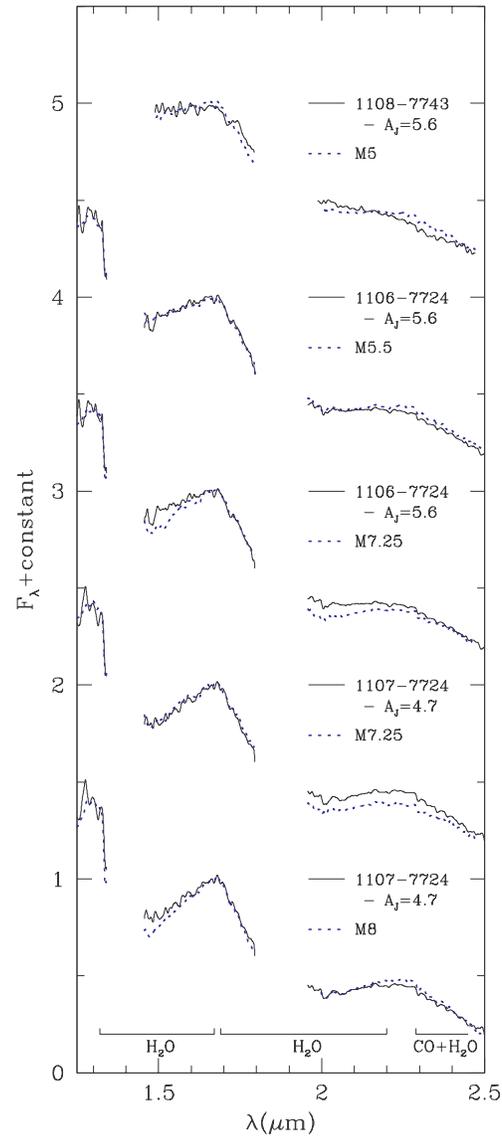}
\caption{
Near-IR spectra of three new members of Chamaeleon~I, 
2M~J1108$-$7743, 2M~J1106$-$7724, and 2M~J1107$-$7724,
compared to data for previously known members with optical spectral types,
T50 (M5), Hn~12W (M5.5), Cha~H$\alpha$~11 (M7.25), and CHSM~17173 (M8).
The spectra of the new members have been dereddened to match the slopes
of the young optical standards.
The spectra are displayed at a resolution of $R=200$ and are normalized at
1.68~\micron.
}
\label{fig:ir}
\end{figure}

To measure the spectral types of 2M~1108$-$7743, 2M~1106$-$7724, and 
2M~1107$-$7724, 
we have compared their H$_2$O and atomic absorption features to those of the 
optically-classified members of Chamaeleon~I that we observed with GNIRS. 
To facilitate this comparison, we dereddened the spectra of the new members
to give them the same slopes as the known members. We also smoothed the spectra
to a low resolution when comparing the broad H$_2$O absorption bands. 
The dereddened, smoothed versions of the spectra are shown in 
Figure~\ref{fig:ir}. 
Each object is compared to the young standards that provide the closest
match to the strength of the H$_2$O absorption. Through this comparison
and a similar analysis of the atomic absorption lines, we classify 
2M~1108$-$7743, 2M~1106$-$7724, and 2M~1107$-$7724 as M4$\pm$1, M6$\pm$1, and 
M7.5$\pm$1, respectively. 
These spectral types correspond to masses of $\sim0.3$, 0.1, and 0.04~$M_\odot$,
respectively, according to the theoretical evolution models of \citet{bar98} 
and \citet{cha00} and the temperature scale of \citet{luh03ic} for ages of 
a few Myr.

\begin{figure}
\epsscale{0.8}
\plotone{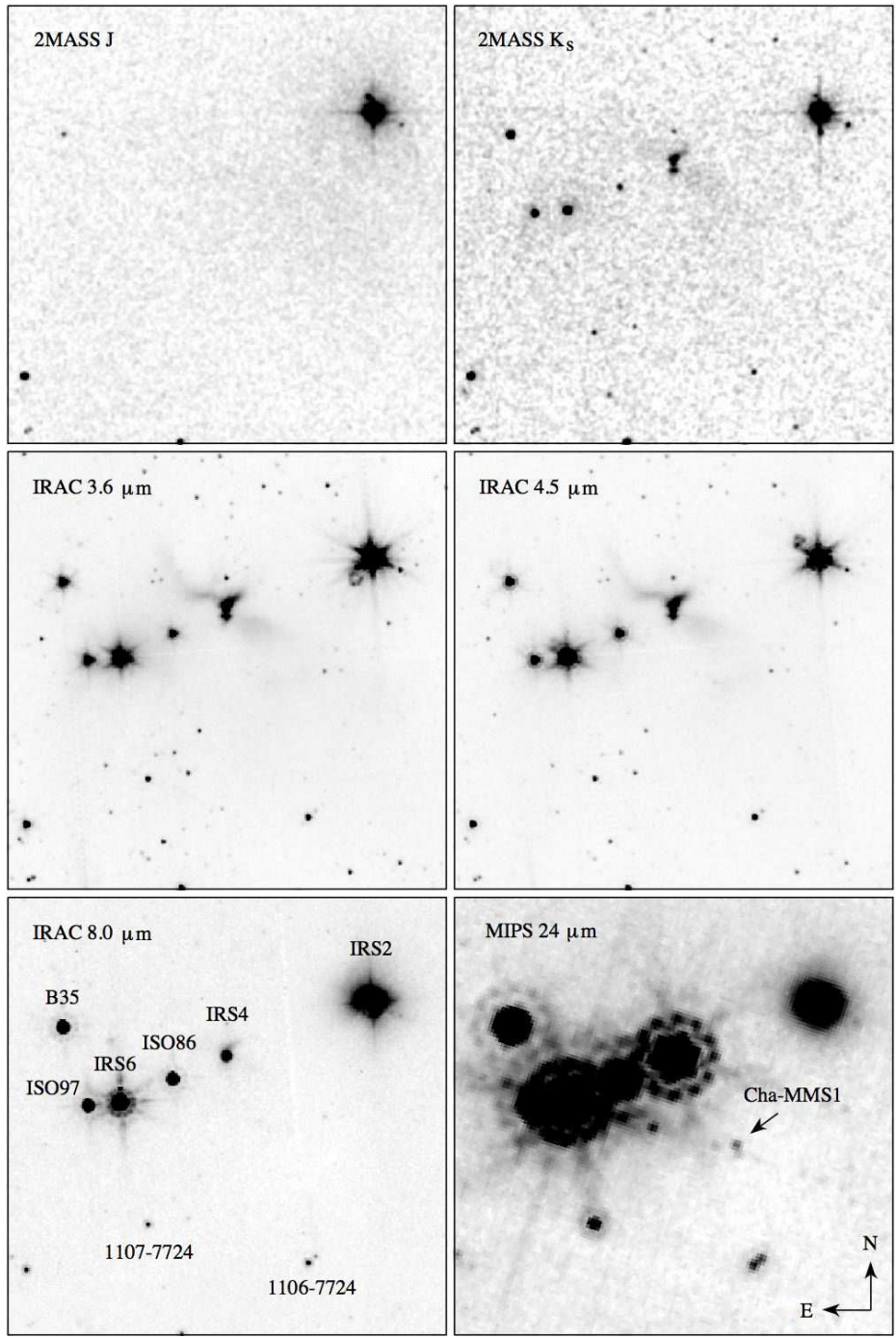}
\caption{
2MASS and {\it Spitzer} images of the region surrounding 
the Cederblad~110 reflection nebula in Chamaeleon~I ($5\arcmin\times5\arcmin$). 
2M~1106$-$7724 and 2M~1107$-$7724 are highly reddened, low-mass disk-bearing
sources discovered in this work. A redder and fainter candidate member
of Chamaeleon~I appears $7\arcsec$ northwest of 2M~1106$-$7724 
(Table~\ref{tab:cand}). The protostar Cha-MMS1 is detected at 24~\micron.
The faint source in the middle of the 24~\micron\ image (southeast of 
Ced~110-IRS4) is a latent image from Ced~110-IRS6. 
}
\label{fig:image}
\end{figure}

The process of dereddening the near-IR spectra of 2M~1108$-$7743, 
2M~1106$-$7724, 
and 2M~1107$-$7724 to match the young standards, which have little if any
extinction \citep{luh07cha}, produces extinction estimates of 
$A_J=5.6$, 5.6, and 4.7, respectively. It is not surprising that these 
objects are highly obscured given their locations. 2M~1108$-$7743 is
only $14\arcsec$ from the protostar IRN while the other two sources
are within the group of embedded stars near Cederblad~110. 
{\it Spitzer} and 2MASS images of 2M~1106$-$7724, 2M~1107$-$7724, and
the protostars in Cederblad~110 are shown in Figure~\ref{fig:image}.
Based on their high levels of extinction and their close proximity to known 
protostars, 2M~1108$-$7743, 2M~1106$-$7724, and 2M~1107$-$7724 may be the 
youngest known low-mass members of the cluster. By combining our extinction 
estimates for these sources with their $H$ and $K_s$ magnitudes, 
the appropriate bolometric corrections, and the distance of Chamaeleon~I, 
we arrive at luminosities of 0.053, 0.035, and 0.020~$L_\odot$,
respectively. Because the $H$- and $K$-band spectra for these sources
do not show any evidence of continuum veiling, these luminosity estimates 
should be uncontaminated by disk emission.

The spectral types, extinctions, luminosities, membership evidence, and 
near-IR photometry for the four new members of Chamaeleon~I are provided in 
Table~\ref{tab:new}. The IRAC and MIPS photometry for these sources are 
presented in the tabulation of {\it Spitzer} data for all known members 
of the cluster in \S~\ref{sec:phot}. The IRAC and MIPS measurements for
the 13 nonmembers are listed in Table~\ref{tab:non}.

\subsection{Remaining Candidates}
\label{sec:cand}

In the previous section, we obtained spectra for a small sample of 
the candidate disk-bearing members of Chamaeleon~I that were
identified in \S~\ref{sec:select} with Figure~\ref{fig:1234}. 
To investigate the nature of the remaining candidates,  
we compare them to the known cluster members in {\it Spitzer} color-magnitude 
and color-color diagrams in Figure~\ref{fig:14}.
Many of the candidates are as red as the reddest known members, indicating
that they are protostellar if they are members.
In addition, bona fide members are likely to be brown dwarfs given that 
nearly all of the candidates are fainter than the stellar members of the 
cluster, although stars with edge-on disks can also appear very faint. 
For instance, Cha~J11081938$-$7731522 has a stellar mass according to 
its spectral type, but it is the faintest and reddest known member in [3.6] 
and $[3.6]-[24]$, respectively, which is explained by the presence of an
edge-on disk \citep{luh07cha}. However, rather than brown dwarfs or stars 
with edge-on disks, the vast majority of the candidates are probably galaxies.

We can refine our sample of candidate members and remove some of the galaxy
contaminants by applying additional criteria.
The initial sample of candidates was defined by colors of $[3.6]-[4.5]>0.15$ 
and $[5.8]-[8.0]>0.3$. However, the sources with $[3.6]-[4.5]\lesssim0.7$ and
$[5.8]-[8.0]\gtrsim1.2$ within that sample have colors that differ
significantly from those of all known members of Chamaeleon~I,
as shown in Figure~\ref{fig:1234}. These objects can be rejected as 
likely galaxies. Because galaxy contamination increases with fainter
magnitudes, brighter candidates are less likely to be galaxies. 
The probability of membership is also higher for candidates that are 
close to known members of Chamaeleon~I. By applying these criteria,
we have identified the most promising candidate disk-bearing members,
which are listed in Table~\ref{tab:cand}.
The brightest candidate, 2MASS~J11085367$-$7521359, 
is probably a solar-mass star if it is associated with Chamaeleon~I.
The magnitudes of the remaining candidates are indicative of brown dwarfs.
As noted in Table~\ref{tab:cand}, some of the candidates may be 
class~I sources or companions to known members.
The new member 2M~1106$-$7724 and the candidate Cha~J11062788$-$7724543 
could comprise the youngest low-mass binary system discovered to date.

In addition to revealing new candidate members of Chamaeleon~I, our
mid-IR measurements provide constraints on the membership of 
sources that have been previously cited as possible members.
In a census of Chamaeleon~I, \citet{luh04cha} assigned membership to a few
objects that lacked spectroscopic classification but that exhibited 
mid-IR excess emission. One of these objects is ISO~13 \citep{nor96,per00}.
The near-IR counterpart for ISO~13 was not positively identified in
previous studies because of uncertainties in the coordinates measured
with {\it ISO}. Based on a comparison of the photometry from {\it ISO}
to measurements from IRAC for sources in the vicinity of the {\it ISO}
coordinates, we conclude that ISO~13 corresponds to 2MASS~J11025579$-$7724304. 
For this source, we measure magnitudes of 13.87, 13.68, 12.68, 9.87, and 
6.54 at 3.6, 4.5, 5.8, 8.0, and 24~\micron, respectively.
When we combine these data with photometry from 
the Two-Micron All-Sky Survey \citep[2MASS,][]{skr06}, we find that the
colors of ISO~13 are relatively blue at 1-5~\micron\ but become very red
at 8 and 24~\micron. These colors are inconsistent with those of 
known members of Chamaeleon~I, as illustrated by the position
of ISO~13 in the IRAC color-color diagram in Figure~\ref{fig:1234}. 
Therefore, we remove ISO~13 from our list of members of Chamaeleon~I because
it is probably a galaxy.
\citet{luh04cha} included ISO~206 in a list of candidate members
based on an apparent excess at 6.7~\micron\ in photometry from {\it ISO}.
However, mid-IR excess emission is not present in the IRAC data for this
source. As a result, there is no longer any reason to consider ISO~206 as
a possible member. 
Finally, \citet{per01} identified three candidate class~I brown dwarfs 
in the vicinity of Cederblad~110 through the presence of near-IR excess 
emission. In our IRAC data, we find that one of these candidates, NIR72, 
is a patch of nebulosity rather than a stellar source while the other 
two objects, NIR84 and NIR89, do not exhibit excess emission, and thus
are not class~I sources.

\subsection{Completeness for Members with Disks}
\label{sec:complete}

\begin{figure}
\epsscale{0.6}
\plotone{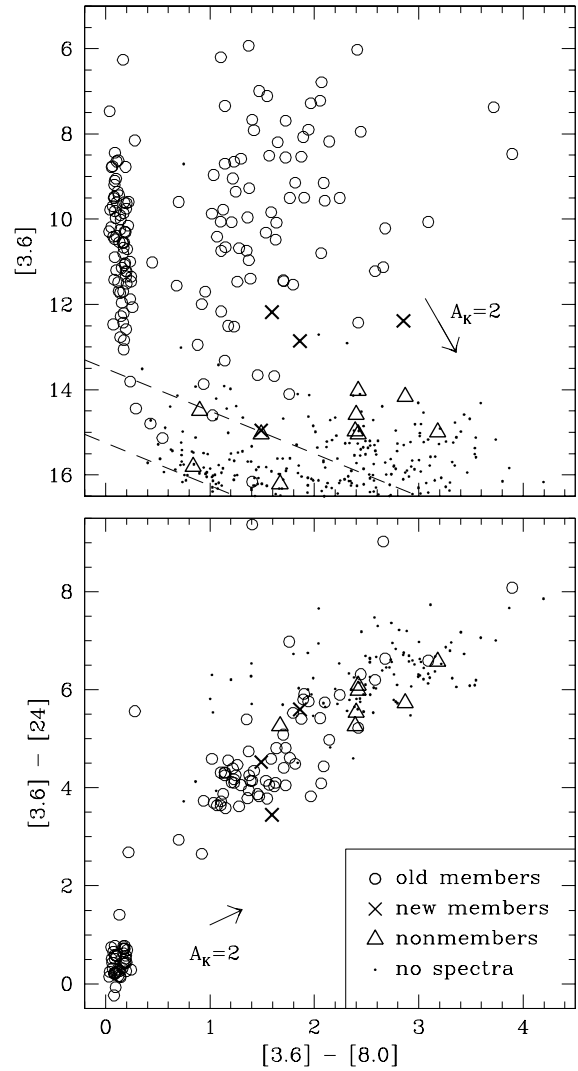}
\caption{
{\it Spitzer} color-magnitude and color-color diagrams for the survey fields 
in the Chamaeleon~I star-forming region that are indicated in 
Figure~\ref{fig:map}. We show the previously known members of Chamaeleon~I 
({\it circles}) and objects that have red colors in Figure~\ref{fig:1234} and
that have been spectroscopically classified as new members
({\it crosses}, Table~\ref{tab:new}) and nonmembers ({\it triangles},
Table~\ref{tab:non}). We also include the remaining
sources from Figure~\ref{fig:1234} that have red colors
($[3.6]-[4.5]>0.15$, $[5.8]-[8.0]>0.3$) and lack spectroscopy ({\it points}), 
most of which are probably galaxies based on their faint magnitudes.
Only objects with photometric errors less than 0.1~mag in all four bands
are shown in these diagrams, with the exception of the new member 
2M~J1108$-$7743.
The 8~\micron\ completeness limits for the shortest and longest exposures 
are shown in the color-magnitude diagram ({\it dashed lines}, 
Fig.~\ref{fig:limit}). The reddening vectors are based on the extinction 
law from \citet{fla07}.
}
\label{fig:14}
\end{figure}

We now use our mid-IR data to examine the completeness of the current 
census of disk-bearing members of Chamaeleon~I.
We consider only the areas that have been imaged in all four bands of 
IRAC, which encompass 170 of the 229 known members of the cluster.  
Within these areas, 156 members have photometric errors less than 0.1~mag
in all IRAC bands, while the remaining 14 known members lack accurate
photometry in at least one band because they are saturated, extended, too 
close to a brighter star, or below the detection limit of all four bands
(Cha-MMS1).
Thus, the photometric criteria that we used for selecting candidate members
with disks in \S~\ref{sec:select} recover all known members with disks 
in the areas with 4-band coverage with the exception of the brightest stars,
faint companions, and the youngest protostars.
As a result, our sample of candidates should
be complete for disk-bearing members in the same range of magnitudes and
extinctions exhibited by the known stellar population.
To quantify the completeness limits of this candidate sample, we have plotted
the photometric completeness limits at 8~\micron\ for the shortest and longest
exposures in the color-magnitude diagram in Figure~\ref{fig:14} 
(\S~\ref{sec:reduction}).
Meanwhile, to evaluate the completeness of the current census of confirmed
members with disks, we consider the remaining candidates that lack 
spectroscopy. As shown in Figure~\ref{fig:14}, only one candidate is
present in the magnitude range of young stars in Chamaeleon~I 
($[3.6]\lesssim12$), which is in Table~\ref{tab:cand}.
Thus, the current census is nearly complete for stars with disks in the fields
imaged by all four bands of IRAC, which encompass most of the Chamaeleon~I 
cloud (Figure~\ref{fig:map}). Our survey is not complete for brown dwarfs
with disks, as indicated by the large number of candidates lacking 
spectroscopy in Figure~\ref{fig:14}. However, based on optical and 
near-IR surveys, the census of substellar cluster members (both with and
without disks) is complete for specific fields and ranges of mass and 
extinction that are described by \citet{luh07cha}.

\section{Global Properties of Disk Population}
\label{sec:global}

\subsection{{\it Spitzer} Photometry for Known Members}
\label{sec:phot}

We wish to use our mid-IR photometry from {\it Spitzer}
to investigate the properties of the disk population in Chamaeleon~I.
To do this, we begin by presenting a tabulation of these measurements
for all known members of the cluster.
The latest census of Chamaeleon~I from \citet{luh07cha} contained 226 sources.
ISO~13 was considered a cluster member in that study, but we now classify it
is as a probable galaxy (\S~\ref{sec:cand}). 
The census from \citet{luh07cha} did not include protostellar sources
that have been detected only at far-IR wavelengths and longward, such as 
Cha-MMS1 \citep{rei96}. Because this source is detected in our 
24~\micron\ data (see Figure~\ref{fig:image})\footnote{The detection
of Cha-MMS1 in the 24~\micron\ data was previously reported by \citet{bel06}.},
we include it in the list of members considered in this work.
Eight young stars from \citet{cov97} are within the area in which 
we have reduced nearly all available IRAC and MIPS data, which has a radius
of $3\arcdeg$ centered at $\alpha=11^{\rm h}07^{\rm m}00^{\rm s}$, 
$\delta=-77\arcdeg10\arcmin00\arcsec$ (J2000).
Three of these stars, RX~J1108.8$-$7519A and B and RX~J1129.2$-$7546, 
have proper motions that are similar to those of known members of Chamaeleon~I, 
while the other five stars are probably members of the
$\epsilon$~Cha association according to their proper motions (see Appendix).
We also find that four of the young stars from \citet{luh07cha}
have proper motions that favor membership in $\epsilon$~Cha rather than
Chamaeleon~I. Therefore, we deduct ISO~13 and these four stars from the 
census of \citet{luh07cha} and add Cha-MMS1, RX~J1108.8$-$7519A and B, 
and RX~J1129.2$-$7546.
When we include the four new members that we have discovered (\S~\ref{sec:new}),
we arrive at census of Chamaeleon~I that contains 229 sources.

The IRAC images encompass 198 of the 229 known members, all of which are
detected in at least one IRAC band, except for the protostar Cha-MMS1 and 
six objects that are unresolved from brighter stars (ESO~H$\alpha$~281, 
2MASS~J11011926$-$7732383B, 2MASS~J11072022$-$7738111, CHXR~73B, T39B, T33B).
The MIPS 24~\micron\ images contain 199 members, 152 of which are detected. 
Among the MIPS detections, seven stars are saturated and one
source, T41, is not measured because it is extended. 
Thus, we have measured IRAC and MIPS photometry for 191 and 144 known
members, respectively. These data are presented in Table~\ref{tab:mem}.
Nondetections at 24~\micron\ are indicated for
members with IRAC measurements. Sources that lack measurements in both
IRAC and MIPS are not included in Table~\ref{tab:mem}, which consist of six
unresolved companions (2MASS~J11011926$-$7732383B, 2MASS~J11072022$-$7738111, 
CHXR~73B, T39B, T33B, CHXR~68B), one member not detected by MIPS and outside
of IRAC (2MASS~J11155827$-$7729046), and 19 members outside of both the
IRAC and MIPS images. ESO~H$\alpha$~281 is unresolved from a brighter 
background star in 2MASS \citep{luh07cha} but is comparable to the latter at 
8~\micron\ and is the dominant source at 24~\micron. Thus, we were able to 
measure photometry for ESO~H$\alpha$~281 at 24~\micron\ but not in the IRAC 
bands.

\begin{figure}
\epsscale{1}
\plotone{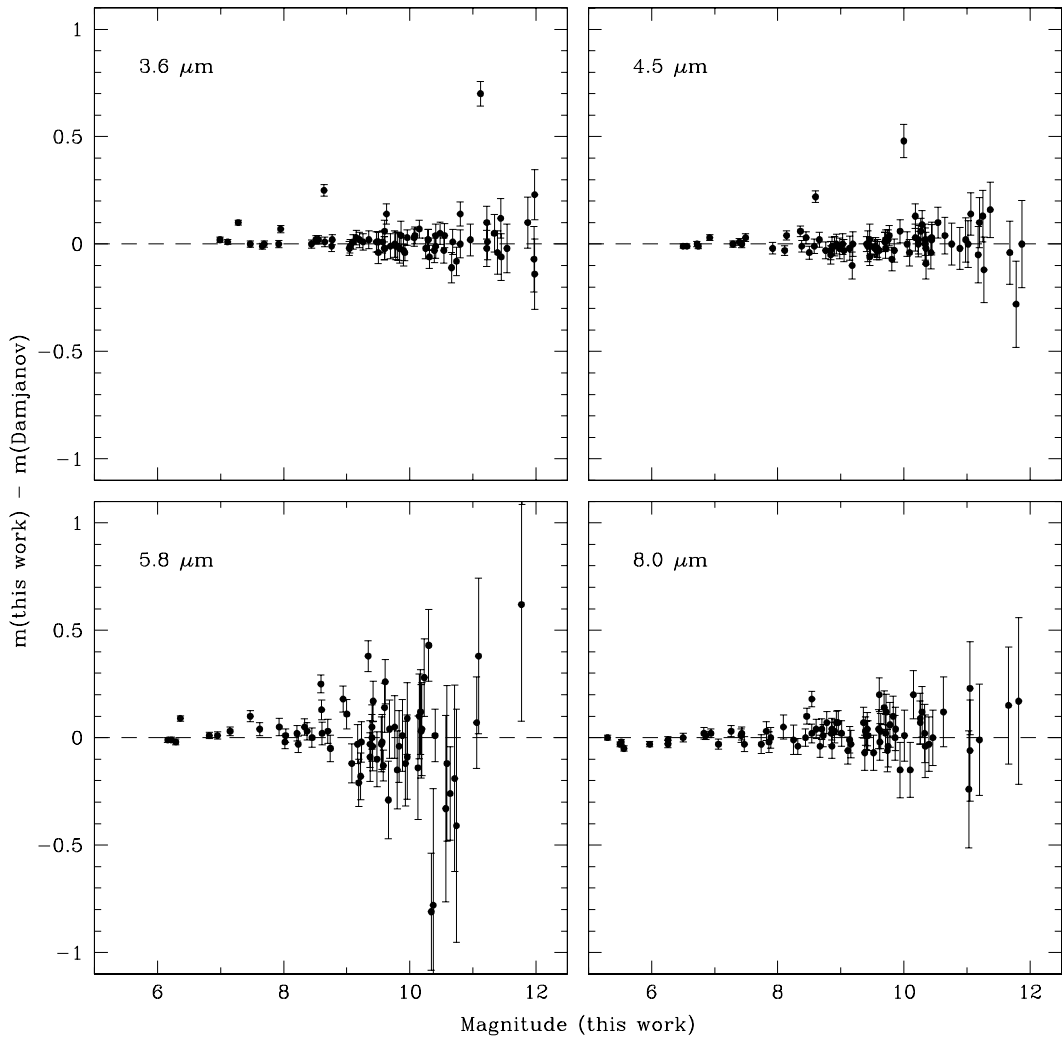}
\caption{
Comparison of the IRAC magnitudes from \citet{dam07} 
to our measurements of the same objects from the same images considered
in that study. The errors reported by \citet{dam07} are indicated.
Nearly all of our measurements have errors of 0.02-0.04~mag.
}
\label{fig:dam1}
\end{figure}

\begin{figure}
\epsscale{0.6}
\plotone{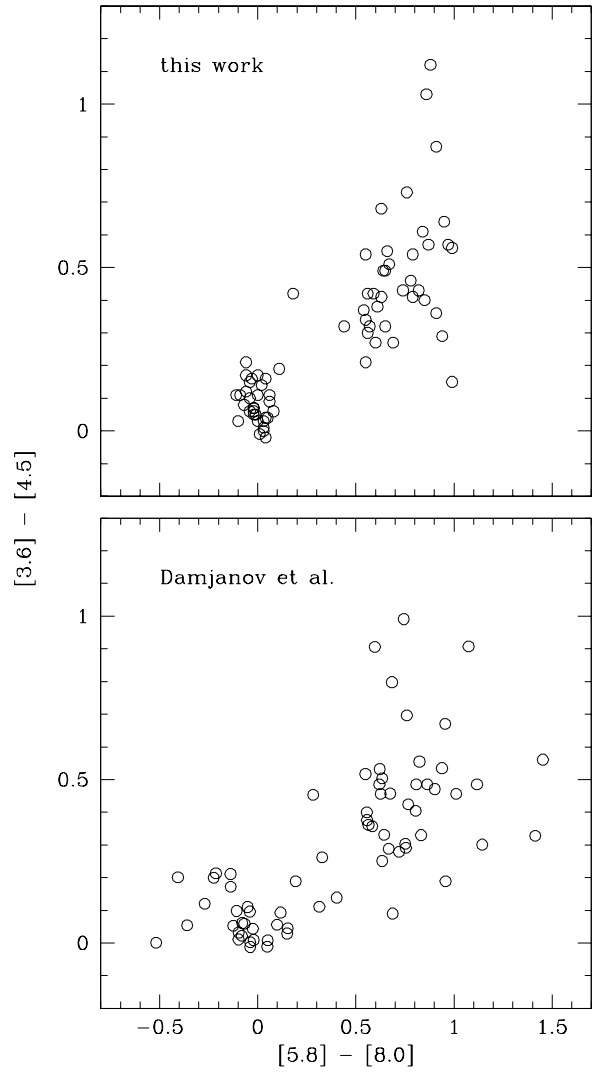}
\caption{
Comparison of the IRAC colors from \citet{dam07} ({\it bottom})
to our measurements of the same objects from the same images considered
in that study ({\it top}).
}
\label{fig:dam2}
\end{figure}

As mentioned in \S~\ref{sec:intro}, previous studies have measured
photometry for members of Chamaeleon~I from the {\it Spitzer} images
considered in this work. \citet{luh05ots,luh05frac,luh05cha} reported
IRAC photometry for known members with spectral types later than M6. 
Those measurements are now superseded by the ones in Table~\ref{tab:mem}, 
which were performed with newer versions of the image processing software
described in \S~\ref{sec:reduction}.
More recently, \citet{dam07} measured IRAC photometry for 81 members from 
the images for AORs 3960320 and 3651328 and MIPS 24~\micron\ photometry 
for 59 members from the images for AORs 3661312 and 3962112. 
To compare the IRAC photometry from \citet{dam07} to our measurements from
the same images, we plot the differences of the IRAC magnitudes from the 
two studies as a function of magnitude in Figure~\ref{fig:dam1}.
The two sets of photometry differ by more than 0.1~mag in many cases,
particularly at fainter levels and at 5.8~\micron. Most of these differences
are within the errors quoted by \citet{dam07}, but those errors are much 
larger than the ones that we estimate for our photometry.  
We find errors of only a few percent for all of our measurements in 
Figure~\ref{fig:dam1}, while the errors from \citet{dam07} are as large as
$\sim50$\%. To illustrate the relative accuracies of these measurements 
without making use of the quoted errors, we compare the IRAC color-color 
diagrams that are produced by the two studies in Figure~\ref{fig:dam2}.
Our colors form two distinct groups, a tight cluster near the origin and 
a broader distribution of significantly redder colors, which are indicative
of stellar photospheres and stars with disks, respectively.
In comparison, the colors from \citet{dam07} exhibit a much larger scatter
and extend to significantly negative colors, which are unphysical for the
spectral types in question. Thus, according to both this comparison and the
formal error estimates, our measurements have significantly higher accuracies 
than those from \citet{dam07}.

\subsection{SED Classifications}
\label{sec:sed}

To classify the IR spectral energy distributions (SEDs) of the members
of Chamaeleon~I, we use the spectral slope defined as
$\alpha= d$~log$(\lambda F_\lambda)/d$~log$(\lambda)$ \citep{lw84,adams87}.
This slope is normally measured between $\lambda\sim2$~\micron\ and 
10-20~\micron.
We have measured photometry at 8 and 24~\micron\ for 182 and 144 members, 
respectively. At least one of these bands is available for 197 members.
At short wavelengths, data at 3.6~\micron\ are available for 164 members. 
Slopes from 3.6~\micron\ to 8 or 24~\micron\ can be computed for 160 members. 
Photometry at $K_s$ (2.2~\micron) is also useful as a short-wavelength
band for spectral slopes since it has been measured for all known members 
\citep[2MASS;][]{luh07cha}.
Therefore, to provide a spectral slope for as many members as possible, we 
have computed slopes between four pairs of bands, 2.2-8, 2.2-24, 3.6-8, and 
3.6-24~\micron. We have dereddened these slopes using the extinctions from
\citet{luh07cha} and the reddening law from \citet{fla07}, except for
10 members that lack spectral types and extinction estimates.
As in \S~\ref{sec:select}, we have adopted the average measurement 
in a given band if an object has been observed at multiple epochs.
The resulting values of $\alpha_{2-8}$, $\alpha_{2-24}$, $\alpha_{3.6-8}$,
and $\alpha_{3.6-24}$ are listed in Table~\ref{tab:alpha} for the
196 known members of Chamaeleon~I that have photometry at 8 or 
24~\micron\ and at shorter wavelengths (i.e., excluding Cha-MMS1).

\begin{figure}
\epsscale{1}
\plotone{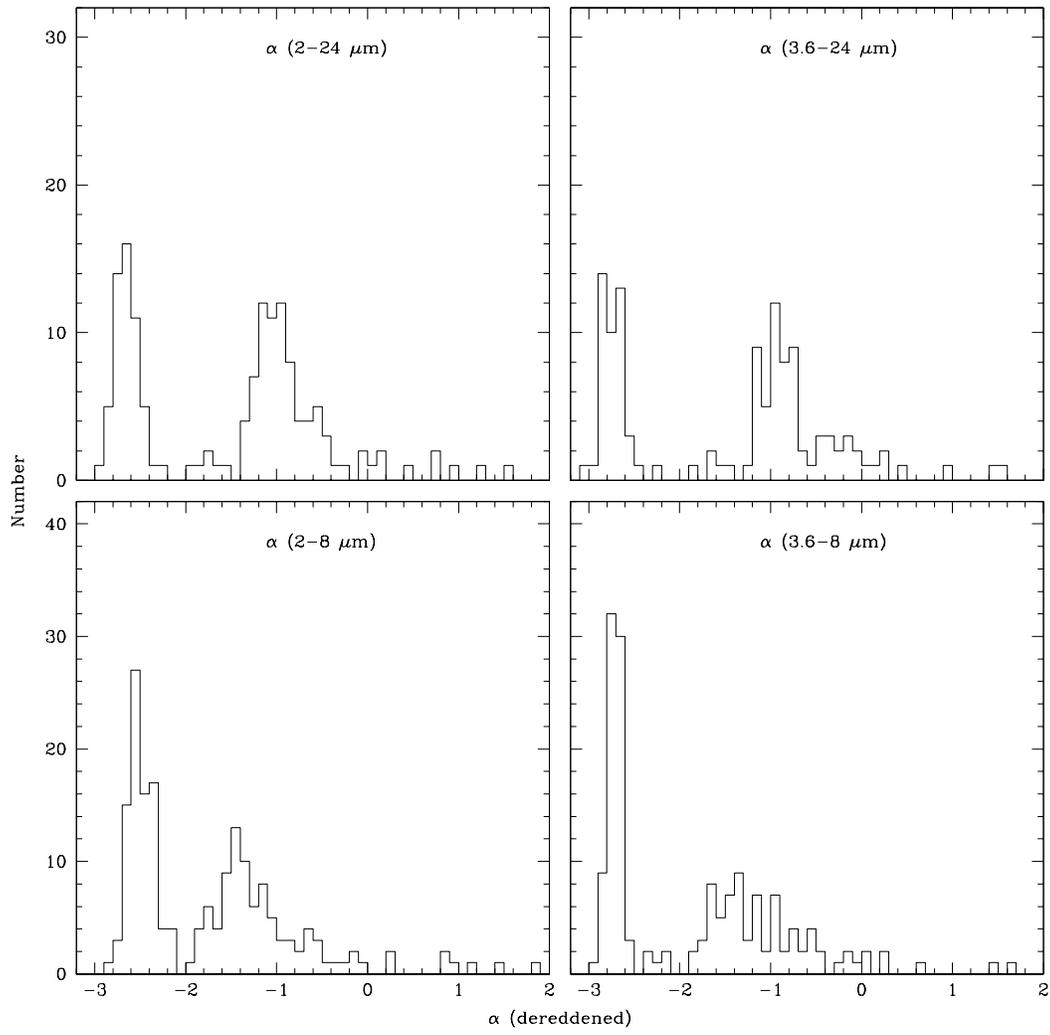}
\caption{
Distributions of spectral slopes for members of Chamaeleon~I
(Table~\ref{tab:alpha}). 
}
\label{fig:alpha2}
\end{figure}

\begin{figure}
\epsscale{1}
\plotone{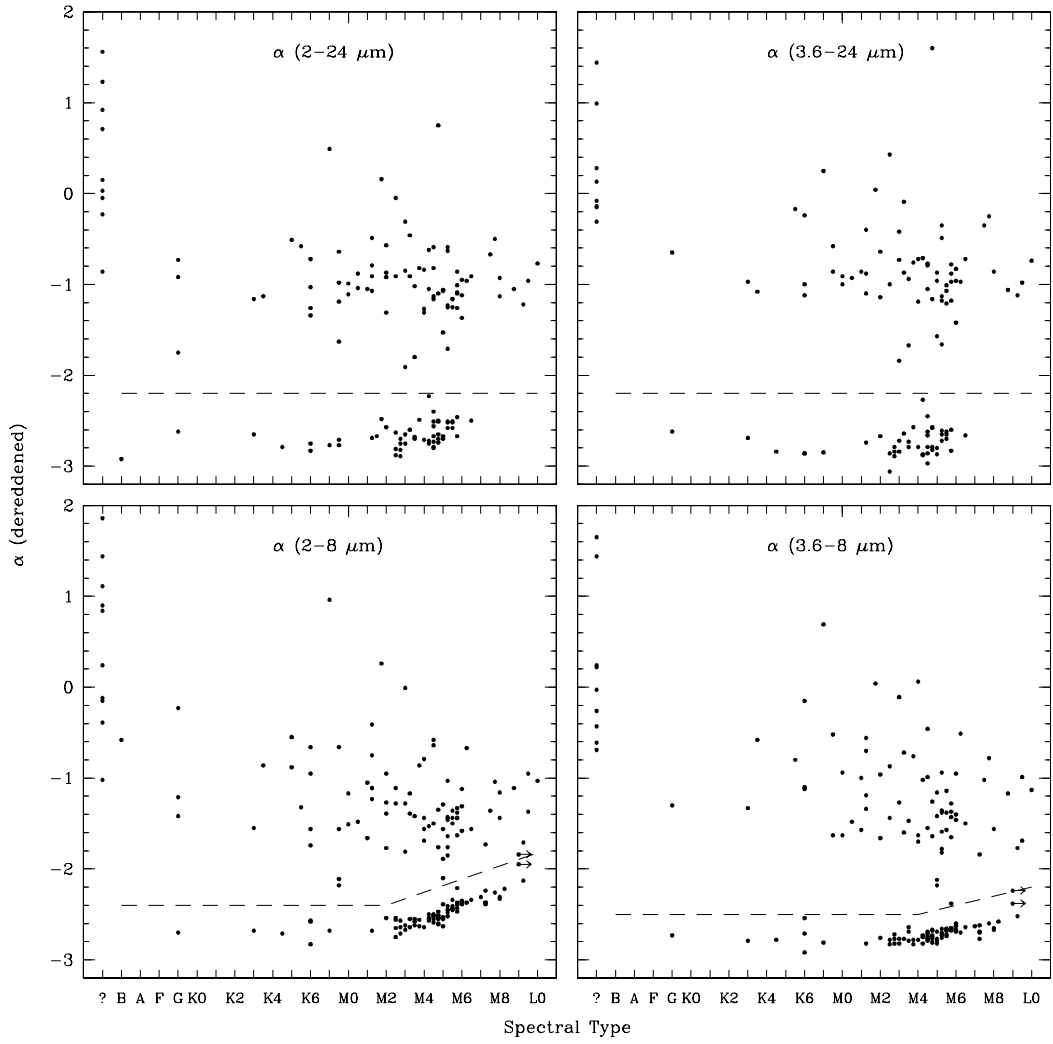}
\caption{
Spectral slopes as function of spectral type for members of Chamaeleon~I 
(Table~\ref{tab:alpha}).
Our adopted boundaries for separating SED classes II and III are indicated
({\it dashed lines}). 
}
\label{fig:alpha1}
\end{figure}

The distributions of spectral slopes from Table~\ref{tab:alpha} are
shown in Figure~\ref{fig:alpha2}.
Like the colors in Figure~\ref{fig:1234}, the slopes form 
two distinct populations that are well-separated from each other. 
The narrow group of bluer objects corresponds to stellar photospheres
while the broader distribution of redder sources represents stars with disks.
We investigate these data in greater detail by plotting the slopes
as a function of spectral type in Figure~\ref{fig:alpha1}.
The widths of the photospheric sequences for $\alpha_{2-24}$ and 
$\alpha_{3.6-24}$ increase at later spectral types because of the larger
photometric errors at 24~\micron\ for cooler, fainter objects.
Most photospheres later than M6 are below the detection limit
at 24~\micron. Because of the close proximity and low background of 
Chamaeleon~I, this detection limit of M6 for stellar photospheres is 
significantly cooler than the limits typically achieved in 
MIPS 24~\micron\ surveys of other star-forming regions.
In the two slopes based on 8~\micron, the separations between the 
two populations are smaller than in $\alpha_{2-24}$ and $\alpha_{3.6-24}$, 
but the photospheric sequences are also tighter
because of the higher photometric accuracies at 8~\micron\ relative to
24~\micron. The distributions of dereddened slopes in Figure~\ref{fig:alpha1} 
are nearly identical to the distributions of observed values except that
the sequences of stellar photospheres are slightly tighter after dereddening.

In Figure~\ref{fig:alpha1}, $\alpha_{2-8}$ and $\alpha_{3.6-8}$ are
approximately constant for stellar photospheres with types of B through K, 
but they change noticeably at later spectral types.
As a result, we should not use fixed thresholds in $\alpha_{2-8}$ and
$\alpha_{3.6-8}$ for identifying stars with disks.
Instead, our adopted thresholds are fits to the photospheric sequences that 
have been offset by 0.3 so that they are significantly greater than the scatter
in each sequence. 
A constant threshold of $-2.2$ is sufficient for identifying
stars with disks for both $\alpha_{2-24}$ and $\alpha_{3.6-24}$,
as shown in Figure~\ref{fig:alpha1}.
Following the standard classification scheme for SEDs of young stars 
\citep{lada87,gre94}, we classify stars with slopes below the thresholds in
Figure~\ref{fig:alpha1} as class~III while redder stars with 
$\alpha<-0.3$, $-0.3\leq\alpha\leq0.3$, and $\alpha>0.3$ are designated as
class~II, flat-spectrum, and class~I, respectively.
We cannot classify Cha-MMS1 with our data because it is not detected at 
wavelengths shorter than 24~\micron, but it is probably in the class~0 stage
or an earlier, prestellar phase \citep{rei96,leh01,bel06}.

The SED classifications produced by $\alpha_{2-8}$, $\alpha_{2-24}$,
$\alpha_{3.6-8}$, and $\alpha_{3.6-24}$ disagree for a small number of sources.
The classes from three of the four slopes agree for T14A, 
C1-2, Ced~110-IRS4, ISO~97, ISO~217, ISO~237, 
Hn10E, C1-25, and Cha~H$\alpha$~1.
For each of these stars, we assign the classification from the three slopes
that agree.
ISO~86 and Ced~110-IRS6 are class~I by $\alpha_{2-8}$ and $\alpha_{2-24}$ 
and are class~II by $\alpha_{3.6-8}$ and $\alpha_{3.6-24}$. We adopt the latter
classification for these stars because of the possibility of variability 
during the time between the $K_s$ and {\it Spitzer} observations.
For the same reason, we adopt the classification from $\alpha_{3.6-8}$ 
over the one from $\alpha_{2-8}$ for T29. 
Because the 8~\micron\ slopes of CHSM~15991 indicate a flat-spectrum class
while the 24~\micron\ slopes are only slightly below the flat/II threshold,
we adopt the former classification for this star. 
The SEDs of T11, CHXR~22E, and CHXR~71 are consistent with stellar photospheres
at $\lambda\leq8$~\micron\ but are much redder at 24~\micron.
SEDs of this kind are indicative of disks with inner holes, which are
also known as transitional disks \citep{cal02,cal05,dal05,esp07a,esp07b,fur07}.
CHXR~76 may have a transitional disk as well, although the size of the 
24~\micron\ excess is small and the star is slightly below the 
adopted II/III thresholds for $\alpha_{2-24}$ and $\alpha_{3.6-24}$.
We classify these four stars as class~II rather than class~III since they 
exhibit disk emission at 24~\micron. 
The SEDs of Cha~J11081938$-$7731522 and ESO~H$\alpha$~569 also become
more steeply rising at longer wavelengths. For these stars, the data 
at 8 and 24~\micron\ indicate classes~II and I, respectively.
Both Cha~J11081938$-$7731522 and ESO~H$\alpha$~569 show evidence of edge-on
disks \citep{luh07cha}, which is consistent with the distinctive shapes of 
their SEDs. Thus, we conclude that they are probably edge-on class~II systems. 
Hn~21E and the new member near IRN, 2M~J1108$-$7743, have different
classifications according to $\alpha_{2-8}$ and $\alpha_{3.6-8}$. 
For these two stars, we assign the classifications from $\alpha_{3.6-8}$
because the 3.6 and 8~\micron\ data were measured with the same apertures 
and at the same time. Both stars are too close to brighter stars for useful
constraints on their 24~\micron\ fluxes. 
Our final SED classifications are provided in Table~\ref{tab:alpha}.

Finally, we examine the SEDs of the seven known members of Chamaeleon~I that 
are within the {\it Spitzer} images but have not been classified with
the spectral slopes in Figure~\ref{fig:alpha1} because they lack photometry 
at 8 and 24~\micron. These measurements are unavailable because the
stars are outside of the images, below the detection limits, 
saturated, or extended, as indicated in Table~\ref{tab:mem} for 
T6, T27, T41, CHXR~54, 2MASS~J11112249$-$7745427, and 2MASS~J11080609$-$7739406.
The seventh member, 2MASS~J11155827$-$7729046, is outside of the
IRAC images and is not detected by MIPS, and thus is not present in 
Table~\ref{tab:mem}. The absence of a detection at 24~\micron\ is sufficient
to demonstrate that this star is class~III. To classify the other six
objects, we measured spectral slopes from $K_s$ and the available IRAC
data and applied thresholds similar to those developed in 
Figure~\ref{fig:alpha1}. We find that T6, T27, and 2MASS~J11112249$-$7745427
are class~II and T41, CHXR~54, and 2MASS~J11080609$-$7739406 are class~III.

Table~\ref{tab:class} summarizes the SED classifications for the known 
members of Chamaeleon~I that are within the {\it Spitzer} images, 
with the exception of the protostar Cha-MMS1.

\subsection{Disk Fraction}

We can use our SED classifications to measure the fraction of stars
and brown dwarfs in Chamaeleon~I that harbor circumstellar disks.
For the disk fraction to be meaningful, it must be measured from 
a sample of cluster members that is unbiased in terms of disks. 
Although some of the early studies of Chamaeleon~I identified members by 
signatures related to disks, recent surveys at X-ray, optical, and 
near-IR wavelengths have been sensitive to both 
class~II and class~III sources \citep{fl04,stel04,luh07cha}. 
The resulting census is unbiased in terms of SED class for most of the cloud 
\citep{luh07cha}. The major exception is at high extinctions, where
class~I and II stars are readily found through mid-IR excesses (such as
three of the new members in this work) while their class~III counterparts 
are much more difficult to identify. However, in a cloud like Chamaeleon~I
that has relatively low extinction ($A_V<4$), any members that are highly 
embedded ($A_V>10$) are probably very young stars in their parent cores, 
and thus are likely to have disks. In other words, it is unlikely that
Chamaeleon~I contains a significant number of heavily obscured class~III
stars. This conclusion is supported by the X-ray surveys of \citet{fl04}
and \citet{stel04}, which were capable of detecting sources of this kind.
Therefore, we include all known members of Chamaeleon~I in our disk
fraction measurement with the exception of the new substellar 
member Cha~1107$-$7626 and the protostar Cha-MMS1. The former is excluded 
because it was found through its disk emission, but it is not within the 
detection limits of previous unbiased surveys. As for Cha-MMS1, 
sources that are less evolved than class~I are generally not
included in disk fraction measurements.

\begin{figure}
\epsscale{0.6}
\plotone{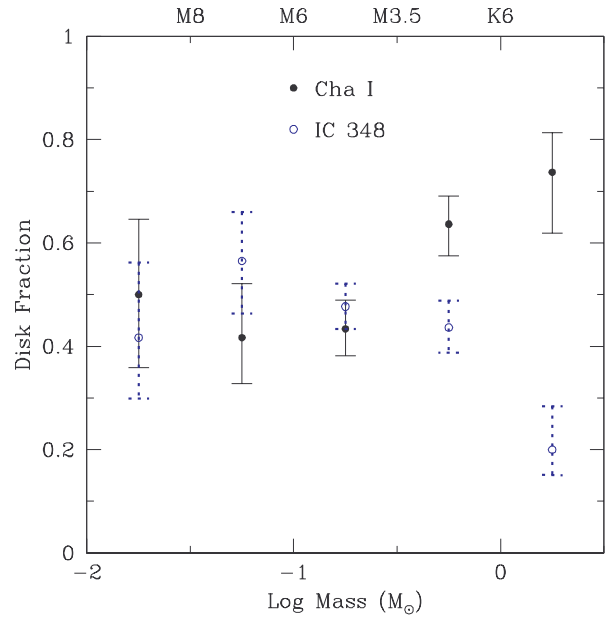}
\caption{
Disk fraction as a function of mass and spectral type for members of 
Chamaeleon~I based on the SED classifications from Figure~\ref{fig:alpha1}.
For comparison, we include the disk fraction of IC~348 computed in the same
manner using data from \citet{luh05frac} and \citet{lada06}.
}
\label{fig:diskfraction}
\end{figure}

We have computed the fraction of members that are class~I or class~II 
as a function of spectral type, which is a good proxy for stellar mass.
We selected the boundaries of the spectral type bins to correspond 
approximately to intervals of 0.5 in the logarithm of mass 
based on the temperature
scale of \citet{luh03ic} and the evolutionary models of \citet{bar98}
and \citet{cha00}. For members that lack spectral classifications, we 
estimate masses and spectral types by combining their photometry, the 
evolutionary models, and an assumed age of 2~Myr, which should be sufficiently 
accurate for identifying the bins in which they belong. 
To compare the disk fractions of Chamaeleon~I and the young cluster 
IC~348, we have applied our SED classification criteria to
the IRAC data for IC~348 from \citet{lada06}. The resulting disk fractions 
as a function of mass for Chamaeleon~I and IC~348 are listed in 
Table~\ref{tab:diskfraction} and are plotted in Figure~\ref{fig:diskfraction}.

\begin{figure}
\epsscale{0.6}
\plotone{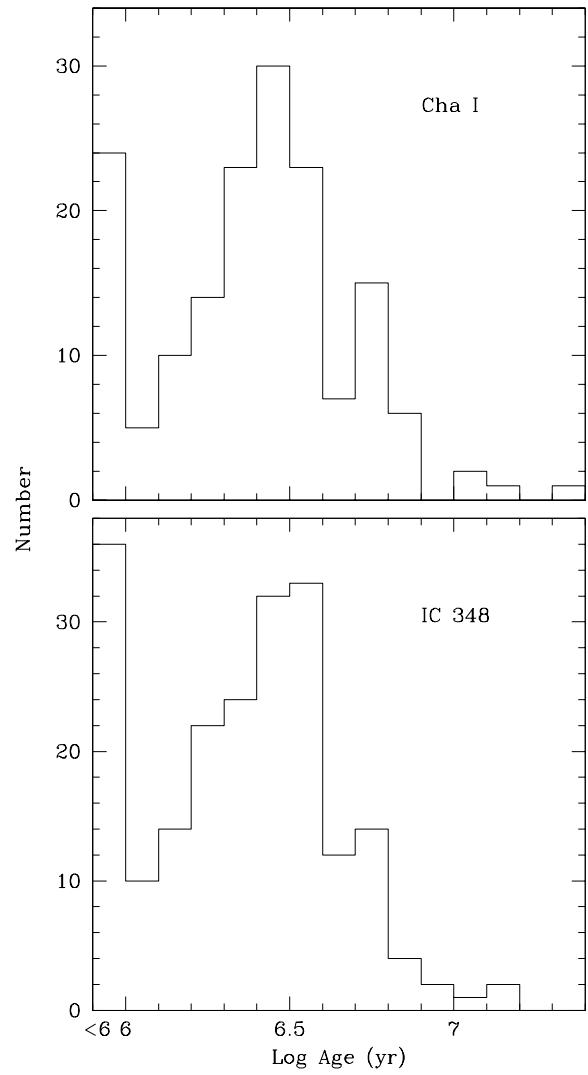}
\caption{
Distributions of isochronal ages for members of Chamaeleon~I ({\it top})
and IC~348 ({\it bottom}) with masses between 0.1 and 1~$M_\odot$
\citep{luh03ic,luh07cha}.
}
\label{fig:ages}
\end{figure}

Through an earlier analysis of a subset of the {\it Spitzer} images,
\citet{luh05frac} found that low-mass stars and brown dwarfs in Chamaeleon~I
exhibit similar disk fractions. We arrive at the same result with our
expanded set of images. Other young clusters like IC~348 and $\sigma$~Ori 
also have disk fractions that are roughly unchanged from low-mass 
stars to brown dwarfs \citep{luh05frac,her07,cab07}. 
As shown in Figure~\ref{fig:diskfraction}, the disk fractions at low masses
in Chamaeleon~I and IC~348 have similar values. 
However, the disk fractions of the two clusters diverge significantly 
above $\sim0.3$~$M_\odot$.
This difference is particularly striking since similar ages have been 
reported for the two clusters \citep{luh03ic,luh07cha}. To compare their
ages in detail, we have estimated the ages of members of 
Chamaeleon~I and IC~348 between 0.1 and 1~$M_\odot$ by combining the
temperatures and luminosities from \citet{luh03ic} and \citet{luh07cha} 
with the evolutionary models of \citet{bar98}. The resulting distributions
of isochronal ages are indistinguishable, as shown in Figure~\ref{fig:ages}.
Thus, the relative disk fractions indicate that the typical lifetime of disks 
around solar-mass stars is shorter in IC~348 than in Chamaeleon~I. 
The higher stellar density of IC~348 relative to Chamaeleon~I is the
only obvious potential source of this difference in disk lifetimes. 
The low-mass stars in IC~348 are segregated toward the outer regions of
the cluster where the density is lower \citep{mue03}, which might explain why 
their disk fraction is similar to that of low-mass stars in Chamaeleon~I. 
Several clusters and associations that are slightly older than 
IC~348 and Chamaeleon~I, such as Upper Sco, NGC~2362, and $\sigma$~Ori 
($\tau\sim5$~Myr), exhibit disk fractions that decrease at higher 
stellar masses in the same manner observed for IC~348 
\citep{car06,dahm07,her07}, which suggests that 
more massive stars have shorter disk lifetimes than low-mass stars in 
most clusters. The solar-mass stars in Chamaeleon~I must have unusually long
disk lifetimes, or their disks must be on the verge of rapidly dissipating.

\subsection{Spatial Distributions of SED Classes}
\label{sec:spatial}

In the previous section, we investigated the dependence of the disk
fraction in Chamaeleon~I on stellar mass. We now consider the 
spatial distribution of disk-bearing stars, which may provide additional
insight into disk evolution.

\begin{figure}
\epsscale{1}
\plotone{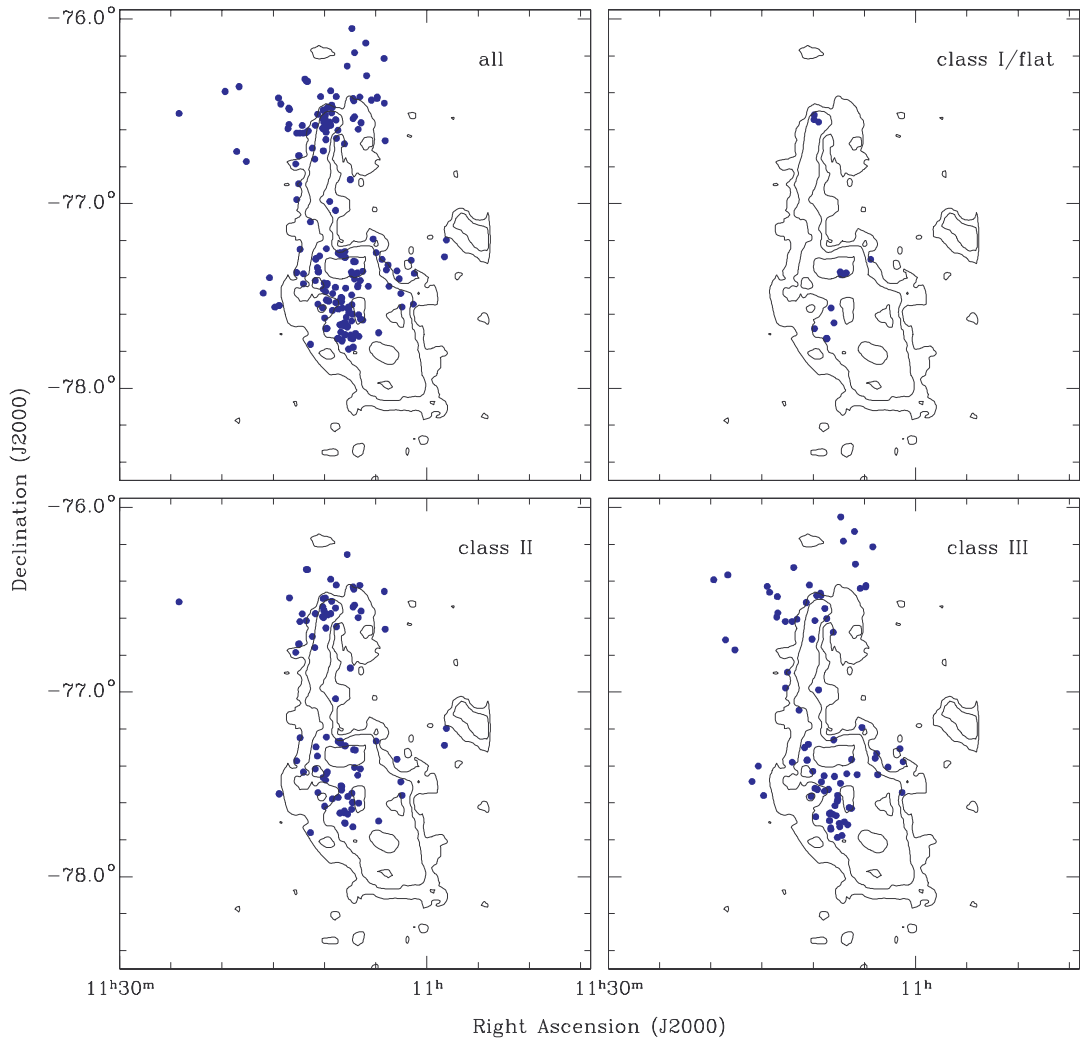}
\caption{
Spatial distributions of members of Chamaeleon~I in the SED classes
from Table~\ref{tab:alpha}.
The contours represent the extinction map of \citet{cam97} at intervals
of $A_J=0.5$, 1, and 2.
}
\label{fig:mapclass}
\end{figure}

As shown in Table~\ref{tab:class}, the southern subcluster exhibits
a slightly higher disk fraction and contains more class~I and flat-spectrum
sources than the northern subcluster, which suggests that the southern 
population is slightly less evolved. A small age difference of this kind is 
also implied by the H-R diagrams of the two subclusters
and the relative levels of extinction \citep{luh07cha}.
The spatial distributions of the members of Chamaeleon~I within the
{\it Spitzer} images are compared in the maps in Figure~\ref{fig:mapclass} 
according to their SED classes. 
Because the protostellar population of Chamaeleon~I is smaller than ones
in younger and richer star-forming regions, the distribution of class~I 
and flat-spectrum sources in Figure~\ref{fig:mapclass} is not particularly 
informative. We simply find that they are highly clustered and are concentrated 
near dense molecular gas, which are typical properties of class~I objects in 
embedded clusters \citep{har02,lada00,lada04,teix06,mue07,all07,gut07,win07}.

The large samples of class~II and III members are more amenable to a 
detailed comparison than the class~I and flat-spectrum objects. 
In Figure~\ref{fig:mapclass}, we find that the
distributions of these two classes are indistinguishable in the southern
subcluster, but differ significantly in the northern one. 
In comparison, \citet{dam07} concluded that stars with disks share the same 
distribution as diskless stars in both subclusters. They did not detect 
the difference in the distributions of classes II and III in the northern
subcluster because their sample of members was three times smaller and
was less complete for class~III objects than the one we are considering. 
The wider class~III distribution in the north can be explained as a 
combination of the segregation of northern low-mass stars to larger radii
\citep{luh07cha} and the lower disk fraction of low-mass stars
relative to solar-type stars (Figure~\ref{fig:diskfraction}).

\subsection{Transitional Disks}

As mentioned in \S~\ref{sec:sed}, transitional disks are characterized
by the presence of large inner holes and gaps. These disks are noteworthy
because their holes may represent evidence of forming giant planets. 
A transitional disk is identified through the shape of its SED, which
exhibits photospheric colors at shorter IR wavelengths followed by a sudden 
onset of excess emission from the disk at longer wavelengths. 
In \S~\ref{sec:sed}, we classified T11, CHXR~22E, CHXR~71, and possibly
CHXR~76 as transitional disks because their spectral slopes showed excess 
emission at 24~\micron\ but not at $\leq8$~\micron. The distinctive SEDs 
of these stars are also illustrated in the color-color diagram 
in Figure~\ref{fig:14}, where they have colors of $[3.6]-[8.0]\sim0.2$ and
$[3.6]-[24]>1$. The transitional disk for T11 (also known as CS~Cha) 
has already been studied in detail through mid-IR spectroscopy with 
{\it Spitzer} \citep{esp07a}.
Meanwhile, a few other cluster members, such as C7-1 and T35, exhibit excess 
emission at both 8 and 24~\micron, but the excess is smaller at 8~\micron. 
In other words, the spectral slopes ending at 8~\micron\ are bluer
than the ones ending at 24~\micron. These stars may possess 
transitional disks, as previously noted by \citet{dam07}. 
Edge-on disks also produce redder slopes at longer wavelengths, as in
Cha~J11081938$-$7731522 and ESO~H$\alpha$~569 (\S~\ref{sec:sed}).
However, unlike a transitional disk, an edge-on disk makes the star
appear much fainter at optical and near-IR wavelengths than unocculted 
stars at the same spectral type, which is not the case for stars 
discussed in this section.

\subsection{Disk Variability}

Our {\it Spitzer} survey has produced high-precision 
mid-IR photometry at multiple epochs for a large fraction of a young
stellar population. As a result, we have a rare opportunity to characterize 
the variability of young stars and brown dwarfs at mid-IR wavelengths. 

\begin{figure}
\epsscale{1}
\plotone{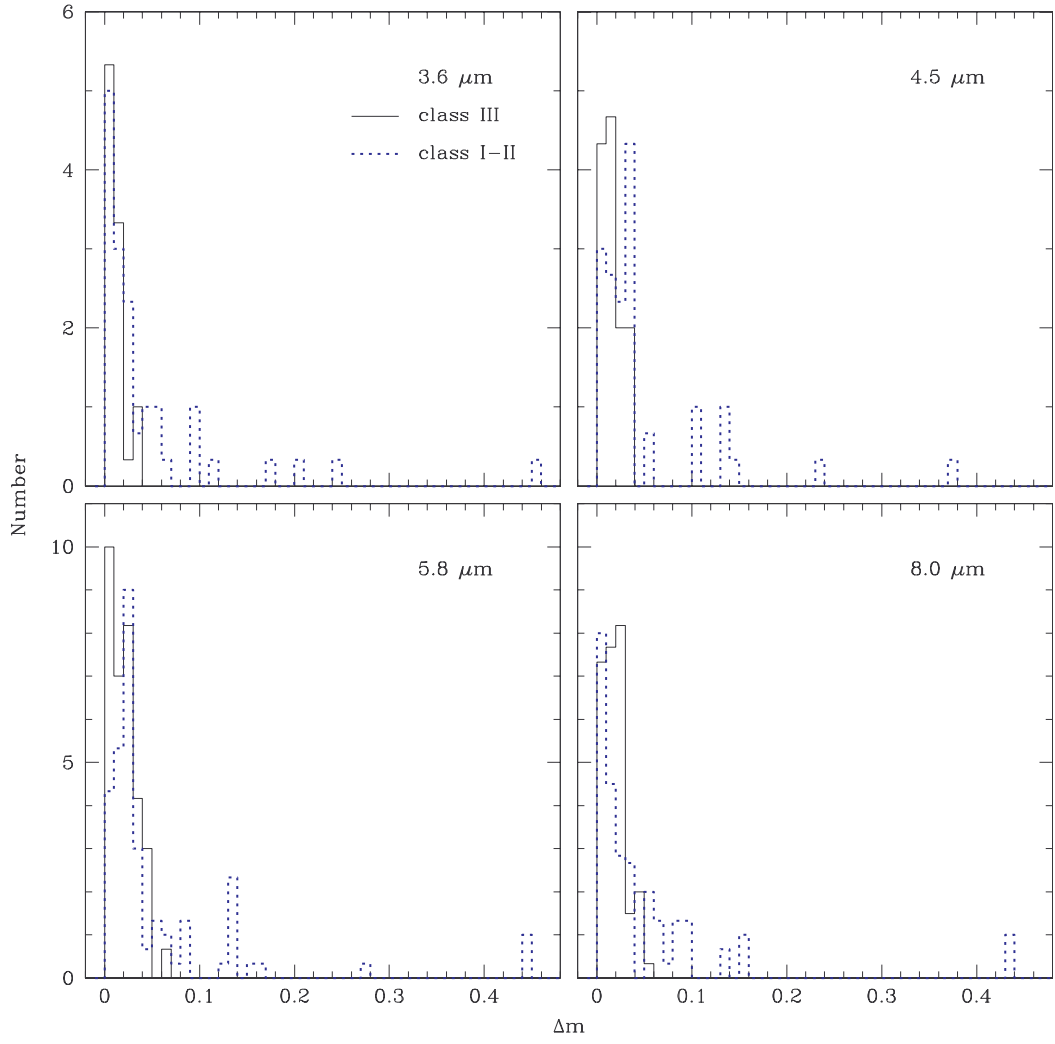}
\caption{
Variability of IRAC magnitudes for members of Chamaeleon~I that have
disks (class~I through class~II, {\it dotted histograms}) and 
members without disks (class~III, {\it solid histograms}).
The histograms represent distributions of differences between a magnitude
and the average magnitude for a given source and band. Each magnitude 
difference is weighted by the inverse of the number of measurements so that 
all members contribute equally.
}
\label{fig:var}
\end{figure}

We have measured photometry at multiple epochs for 27, 30, 70, and 61 members 
at 3.6, 4.5, 5.8, and 8.0~\micron, respectively. The errors are less than 
0.06~mag for 26, 29, 64, and 54 members, respectively.
Fewer members have multiple measurements at shorter wavelengths because 
the saturation limit is fainter at 3.6 and 4.5~\micron.
To quantify the variability of these data, we have computed the 
difference between each magnitude and the average magnitude for a given
object and wavelength. We then constructed histograms of these magnitude 
differences, which are weighted by the inverse of the number 
of measurements so that all members contribute equally to the distribution.
In Figure~\ref{fig:var}, the resulting histograms are shown separately for 
each IRAC band and for stars with and without disks.
In all four bands, nearly all sources with $\Delta m>0.05$~mag have disks,
which indicates that disks are responsible for these large photometric 
variations, either directly through fluctuations in the disk emission
or indirectly through the effect of the disk on the emergent flux from
the stellar photosphere. Based on Figure~\ref{fig:var},
the frequency of large photometric variations appears to be similar among 
the four IRAC bands.
A comparison of this kind between diskless and disk-bearing stars at
24~\micron\ is problematic because of the larger photometric errors,
particularly for the fainter class~III sources.

\section{Conclusions}

We have performed a thorough census of the circumstellar disk population of the 
Chamaeleon~I star-forming region using mid-IR images obtained with the
{\it Spitzer Space Telescope}.
After analyzing most of the images that have been collected in Chamaeleon~I 
with IRAC (3.6-8~\micron) and MIPS (24~\micron) onboard {\it Spitzer}, 
we have searched for new disk-bearing members of the cluster
by identifying sources with red colors in these data. Through spectroscopy
of a small sample of the resulting candidates, we have confirmed the
membership of four objects, which we classify as M4, M6, M7.5, and L0.
The first three sources are located in highly embedded areas ($A_J\sim5$) 
near known protostars, indicating that they may be among the youngest 
low-mass members of the cluster.
The L0 source is the coolest known member of Chamaeleon~I and has a mass 
of 0.004-0.01~$M_\odot$ according to the luminosities predicted by
theoretical evolutionary models \citep{bur97,cha00}, making it the 
least massive brown dwarf for which a disk has been reliably detected.
With the exception of stars with edge-on disks, our survey demonstrates
that the current census of disk-bearing members of Chamaeleon~I is complete
at stellar masses for areas covered by all four bands of IRAC, which 
encompass most of the cloud. At faint magnitudes that correspond to
substellar cluster members, the {\it Spitzer} images detect hundreds of red
sources. Most of these objects are probably galaxies, but we have 
identified a sample of the most promising candidate members, 
which include possible low-mass companions and protostellar sources. 

We have presented a tabulation of IRAC and MIPS photometry for 191 and 144
known members of Chamaeleon~I, respectively. We have used these data to 
classify the SEDs of the 203 members that are encompassed by the IRAC and
MIPS images and that are detected by IRAC (i.e., excluding the protostar 
Cha-MMS1), arriving at 4, 10, 94, and 95 members 
that are class~I, flat spectrum, class~II, and class~III, respectively. 
The disk fraction (I+flat+II/I+flat+II+III) is roughly constant at 
$\sim50$\% from 0.01 to 0.3~$M_\odot$, which closely resembles measurements 
in IC~348. The disk fraction in Chamaeleon~I increases with higher stellar 
masses to a value of $\sim65$\% at 1-2~$M_\odot$, whereas disks in IC~348 are 
much less common in this mass range ($\sim20$\%). 
Because the two clusters have the same 
age, this comparison indicates that solar-type stars in Chamaeleon~I have 
longer disk lifetimes than their counterparts in IC~348. The high stellar
density of the more massive stars in IC~348 may have contributed to 
their shorter disk lifetimes.
We have also examined the spatial distribution of disk-bearing members of
Chamaeleon~I. The distributions of classes II and III are similar in
the southern subcluster, while the class~III sources have a wider  
distribution in the northern one. We interpret the relative distributions
in the north as a reflection of the segregation of low-mass stars to larger 
radii in the northern subcluster \citep{luh07cha} combined with the lower 
disk fraction of low-mass stars relative to members at higher masses. 
Finally, we have identified possible transitional disks around several 
cluster members, and have used our multi-epoch measurements to demonstrate that
stars with disks exhibit greater mid-IR variability than diskless stars.

\acknowledgements

We thank Eric Feigelson for comments on the manuscript
and Gus Muench for helpful discussions.
K.~L. and P.~A. were supported by grant AST-0544588 from the National Science 
Foundation. This publication makes use of data products from 2MASS, which
is a joint project of the University of Massachusetts and the Infrared
Processing and Analysis Center/California Institute of Technology, funded 
by NASA and the NSF.

\appendix

\section{Membership of Young Stars Surrounding the Chamaeleon~I Cloud}
\label{sec:app}

\citet{fri98} demonstrated that the stars in the vicinity of Chamaeleon~I 
consist of two kinematic subgroups: a rich group of young stars
associated with the Chamaeleon~I cloud, and a more dispersed, older group 
with larger motion. The latter has been referred to as the $\epsilon$~Cha 
group \citep{mam00} because it is centered on $\epsilon$~Cha and
the Herbig Ae star HD~104237. The group appears to have an age of 
$\sim6$~Myr and a distance of $d\simeq114$~pc \citep{fei03,luh04eta}.
Proper motions with accuracies of $\sim$5 mas/yr (e.g., UCAC2) are 
sufficient for distinguishing between the two populations.
Therefore, to clarify the membership status of the young stars at distances of 
$\sim1$-3$\arcdeg$ from the center of the Chamaeleon~I cloud 
\citep{cov97,luh07cha}, we compare their proper
motions to those of Chamaeleon~I and the $\epsilon$~Cha group.

The median proper motion for the 61 of the 226 sources in the census of
Chamaeleon~I from \citet{luh07cha} that have UCAC2 proper 
motions \citep{zac04} is $\mu_{\alpha}, \mu_{\delta}= -21$, +2~mas/yr. 
The median proper motion of the members of the $\epsilon$~Cha 
group proposed by \citet{mam00} and \citet{zs04} is 
$\mu_{\alpha}, \mu_{\delta}=-40$, $-$7 mas/yr. The uncertainties in 
these median proper motions are at the level of $\sim$1~mas/yr.
We first consider the eight young stars from \citet{cov97} that are
within $3\arcdeg$ from Chamaeleon~I.
The UCAC2 proper motions of RX~J1108.8$-$7519A and B and RX~J1129.2$-$7546
are consistent with membership in Chamaeleon~I. \citet{mam00} classified
RX~J1158.5$-$7754A (= DW~Cha) as a member of the $\epsilon$~Cha group, 
and we assume that its companion (for which a reliable proper motion
is not available) is a member of that group as well. 
RX~J1150.4$-$7704 was classified as a ``Cha-Near" member by 
\citet[][``Cha-Near" is ostensibly the same as the $\epsilon$ Cha group]{zs04},
and the UCAC2 proper motion corroborates membership in the $\epsilon$~Cha 
group. The UCAC2 proper motion for RX~J1149.8$-$7850 (= DZ~Cha) 
also indicates membership in $\epsilon$~Cha rather than Chamaeleon~I. 
For the final star from \citet{cov97}, RX~J1123.2$-$7924, 
the proper motion from \citet{duc06} is inconsistent with 
both Chamaeleon~I and $\epsilon$ Cha. However, RX~J1123.2$-$7924
was incorrectly matched to 2MASS~J11231052$-$7924434 in \citet{duc06}.
By examining the finder chart from \citet{alc95}, 
we find that RX~J1123.2$-$7924 is instead 2MASS~J11225562$-$7924438.
The UCAC2 proper motion for
this star is $\mu_{\alpha}, \mu_{\delta}=-32$, $-$12 mas/yr 
(errors of 6 mas/yr in each component). Thus, we classify it is a likely 
member of $\epsilon$~Cha. We also have examined the proper motions of 
young stars from \citet{luh07cha} that are between $\epsilon$~Cha
and the Chamaeleon~I cloud.
Using proper motions from UCAC2 and from available optical and IR sky surveys,
we find that 2MASS~J11183572$-$7935548, 2MASS~J11432669$-$7804454, 
2MASS~J11404967$-$7459394, and 2MASS~J11334926$-$7618399 are more likely to
be members of $\epsilon$~Cha than Chamaeleon~I. 
Other young stars from \citet{luh07cha} that are between these two
populations have the proper motions that indicate membership, are 
inconclusive, or are unavailable. 

We present the IRAC and MIPS data, spectral slopes, and SED classifications
for the probable members of the $\epsilon$~Cha association in 
Tables~\ref{tab:eps} and \ref{tab:eps2}.
We note that one of the $\epsilon$~Cha stars has a flat-spectrum SED, 
which is surprising for a population as old as $\epsilon$~Cha 
\citep[$\tau\sim6$~Myr,][]{luh04eta}.

\clearpage

\LongTables

\begin{deluxetable}{rrcclrrl}
\tabletypesize{\scriptsize}
\tablewidth{0pt}
\tablecaption{IRAC Observing Log\label{tab:iraclog}}
\tablehead{
\colhead{} &
\colhead{} &
\colhead{3.6/5.8 Center} &
\colhead{4.5/8.0 Center} &
\colhead{Dimensions} &
\colhead{Angle\tablenotemark{a}} &
\colhead{Exp Time\tablenotemark{b}} &
\colhead{Date} \\
\colhead{AOR} &
\colhead{PID} &
\colhead{(J2000)} &
\colhead{(J2000)} &
\colhead{(arcmin)} &
\colhead{(deg)} &
\colhead{(s)} &
\colhead{(UT)}
}
\startdata
 6526208 &     36 & 11 08 46 $-$77 37 19 & 11 06 45 $-$77 39 37 & 20 $\times$ 15 &  72 &   968 & 2004 Feb 19 \\
 5662720 &    173 & 11 02 59 $-$77 36 25 & 11 01 38 $-$77 31 09 & 13 $\times$ 7 & 142 & 20.8 HDR & 2004 May 1 \\
 5663232 &    173 & 11 24 06 $-$79 27 30 & 11 22 21 $-$79 22 40 & 13 $\times$ 6 & 135 & 20.8 HDR & 2004 May 1 \\
         &        & 11 17 50 $-$80 31 00 & 11 15 56 $-$80 26 09 & 13 $\times$ 6 & 135 & 20.8 HDR &            \\
 5663744 &    173 & 11 50 46 $-$77 07 50 & 11 49 55 $-$77 01 33 & 12 $\times$ 6 & 148 & 20.8 HDR & 2004 May 26 \\
         &        & 12 05 00 $-$77 34 30 & 12 04 04 $-$77 28 30 & 12 $\times$ 6 & 148 & 20.8 HDR &            \\
         &        & 12 17 30 $-$77 56 30 & 12 16 28 $-$77 50 40 & 12 $\times$ 6 & 148 & 20.8 HDR &            \\
 3960320 &     37 & 11 07 44 $-$77 34 46 & 11 07 48 $-$77 28 03 & 33 $\times$ 29 &   3 & 20.8 HDR & 2004 Jun 6 \\
\enddata
\tablenotetext{a}{Position angle of the long axis of the maps.}
\tablenotetext{b}{Total exposure time for each position and filter. 
``HDR" indicates that 0.4~s exposures were also obtained.}
\end{deluxetable}

\begin{deluxetable}{rrclrrl}
\tabletypesize{\scriptsize}
\tablewidth{0pt}
\tablecaption{MIPS 24~\micron\ Observing Log\label{tab:mipslog}}
\tablehead{
\colhead{} &
\colhead{} &
\colhead{Center} &
\colhead{Dimensions} &
\colhead{Angle\tablenotemark{a}} &
\colhead{Exp Time} &
\colhead{Date} \\
\colhead{AOR} &
\colhead{PID} &
\colhead{(J2000)} &
\colhead{(arcmin)} &
\colhead{(deg)} &
\colhead{(s)} &
\colhead{(UT)}
}
\startdata
 5706752 &    173 & 11 09 13 $-$77 29 10 &  8 $\times$ 7 &  55 &   220 & 2004 Feb 3 \\
         &        & 11 12 09 $-$77 21 22 &  8 $\times$ 5 &  55 &   480 &            \\
         &        & 11 07 00 $-$77 32 30 &  7 $\times$ 6 &  55 &   160 &            \\
 4549376 &     72 & 11 58 29 $-$77 54 29 &  8 $\times$ 7 &  61 &   280 & 2004 Feb 22 \\
         &        & 12 01 46 $-$77 47 39 &  8 $\times$ 5 &  61 &   600 &            \\
 5707008 &    173 & 11 10 52 $-$76 44 20 &  8 $\times$ 7 & 115 &   140 & 2004 Apr 8 \\
         &        & 11 14 15 $-$76 48 39 &  8 $\times$ 5 & 115 &   360 &            \\
 9410304 &    139 & 11 02 31 $-$77 39 46 & 17 $\times$ 8 &  28 &    37 & 2004 Apr 8 \\
\enddata
\tablenotetext{a}{Position angle of the long axis of the maps.}
\end{deluxetable}

\begin{deluxetable}{clllll}
\tabletypesize{\scriptsize}
\tablewidth{0pt}
\tablecaption{Spectroscopy Observing Log\label{tab:speclog}}
\tablehead{
\colhead{} &
\colhead{} &
\colhead{} &
\colhead{} &
\colhead{$\lambda$} &
\colhead{} \\
\colhead{Night} &
\colhead{Date} & 
\colhead{Telescope + Instrument} & 
\colhead{Disperser} &
\colhead{($\mu$m)} &
\colhead{$\lambda/\Delta \lambda$}
}
\startdata
1 & 2005 Mar 23 & Gemini + GNIRS & 31.7 cross-dispersed & 1-2.5 & 2000 \\
2 & 2005 Mar 24 & Gemini + GNIRS & 31.7 cross-dispersed & 1-2.5 & 2000 \\
3 & 2006 Feb 9 & Magellan~II + LDSS-3 & VPH red grism & 0.68-1.1 & 1300 \\ 
4 & 2006 Feb 10 & Magellan~II + LDSS-3 & VPH red grism & 0.58-1 & 1300 \\
5 & 2006 Feb 11 & Magellan~II + LDSS-3 & VPH red grism & 0.58-1 & 1300 \\
\enddata
\end{deluxetable}

\clearpage

\begin{deluxetable}{llllllllll}
\tabletypesize{\scriptsize}
\tablewidth{0pt}
\tablecaption{New Members of Chamaeleon I\label{tab:new}}
\tablehead{
\colhead{} &
\colhead{} &
\colhead{$T_{\rm eff}$\tablenotemark{a}} &
\colhead{} &
\colhead{$L_{\rm bol}$} &
\colhead{Membership} &
\colhead{} &
\colhead{} &
\colhead{} &
\colhead{} \\
\colhead{Name} &
\colhead{Spectral Type} &
\colhead{(K)} &
\colhead{$A_J$} &
\colhead{($L_\odot$)} &
\colhead{Evidence\tablenotemark{b}} &
\colhead{$J$\tablenotemark{c}} &
\colhead{$H$\tablenotemark{c}} &
\colhead{$K_s$\tablenotemark{c}} &
\colhead{Night}}
\startdata
2M J11062942$-$7724586 & M6$\pm$1 & 2990 & 5.6 & 0.035 & $A_V$,H$_2$O,ex & \nodata & 15.20$\pm$0.10 & 13.81$\pm$0.06 & 1 \\
2M J11070369$-$7724307 & M7.5$\pm$1 & 2795 & 4.7 & 0.020 & $A_V$,H$_2$O,ex & \nodata & 15.26$\pm$0.12 & 13.95$\pm$0.07 & 1 \\
Cha J11070768$-$7626326 &  L0$\pm$1 & 2200 & 0 & 0.00033 & e,ex & 17.61$\pm$0.03 & 16.80$\pm$0.03 & 15.91$\pm$0.02 & 4,5 \\
2M J11084296$-$7743500 & M4$\pm$1 & 3270 & 5.6 & 0.053 & $A_V$,e,ex & \nodata & 14.87$\pm$0.10 & 13.48$\pm$0.05  & 2 \\
\enddata
\tablenotetext{a}{Temperature scale from \citet{luh03ic}. 
For Cha J11070768$-$7626326, that scale has been extrapolated to L0.}
\tablenotetext{b}{Membership in Chamaeleon is indicated by $A_V\gtrsim1$ and
a position above the main sequence for the distance of Chamaeleon (``$A_V$"),
strong emission lines (``e"), IR excess emission (``ex"),
or the shape of the gravity-sensitive H$_2$O bands (``H$_2$O").}
\tablenotetext{c}{From the 2MASS Point Source Catalog for 2M J11062942$-$7724586 
and 2M J11070369$-$7724307 and from the ISPI images of \citet{luh07cha} for 
Cha J11070768$-$7626326.}
\end{deluxetable}

\begin{deluxetable}{llllll}
\tabletypesize{\scriptsize}
\tablewidth{0pt}
\tablecaption{{\it Spitzer} Photometry for Nonmembers\label{tab:non}}
\tablehead{
\colhead{Name} &
\colhead{[3.6]} &
\colhead{[4.5]} &
\colhead{[5.8]} &
\colhead{[8.0]} &
\colhead{[24]} 
}
\startdata
Cha J11055362$-$7638533 & 14.59$\pm$0.03 & 13.82$\pm$0.03 & 12.90$\pm$0.05 & 12.19$\pm$0.06 &        8.89$\pm$0.06        \\
2M J11081204$-$7622124 &  14.16$\pm$0.03 & 13.73$\pm$0.03 & 13.48$\pm$0.05 & 11.29$\pm$0.04 &      8.31$\pm$0.05         \\
Cha J11081724$-$7631107 & 15.04$\pm$0.03 & 14.78$\pm$0.04 & 14.76$\pm$0.04 & 13.55$\pm$0.04 &      \nodata          \\
Cha J11083029$-$7733067 & 14.49$\pm$0.03 & 14.28$\pm$0.03 & 14.33$\pm$0.04 & 13.59$\pm$0.06 &      \nodata          \\
Cha J11084006$-$7627085 & 15.81$\pm$0.02 & 15.51$\pm$0.02 & 15.58$\pm$0.06 & 14.97$\pm$0.08 &      \nodata          \\
2M J11085651$-$7645154 & 12.47$\pm$0.02 & 12.48$\pm$0.02 & 12.46$\pm$0.05 & 12.34$\pm$0.08 &     \nodata           \\
Cha J11090876$-$7626234 & 16.22$\pm$0.02 & 15.89$\pm$0.02 & 15.64$\pm$0.06 & 14.55$\pm$0.08 &      10.75$\pm$0.17        \\
2M J11105185$-$7642259 & 13.79$\pm$0.02 & 13.77$\pm$0.03 & 13.70$\pm$0.07 & 13.43$\pm$0.11 &     \nodata           \\
2M J11105353$-$7726389 & 12.44$\pm$0.02 & 12.38$\pm$0.02 & 12.33$\pm$0.05 & 12.29$\pm$0.05 &     \nodata           \\
Cha J11112420$-$7637434 & 14.97$\pm$0.03 & 14.20$\pm$0.03 & 13.34$\pm$0.04 & 12.58$\pm$0.05 &     9.58$\pm$0.10           \\
Cha J11114773$-$7642167 &  14.02$\pm$0.02 & 13.19$\pm$0.02 & 12.42$\pm$0.04 & 11.60$\pm$0.05 &        7.90$\pm$0.05        \\
Cha J11120546$-$7631376 & 15.00$\pm$0.02 & 13.89$\pm$0.02 & 12.92$\pm$0.04 & 11.82$\pm$0.04 &      8.28$\pm$0.05          \\
Cha J11130714$-$7638575 &   15.04$\pm$0.03 & 14.20$\pm$0.04 & 13.56$\pm$0.07 & 12.62$\pm$0.06 &      8.79$\pm$0.06          \\
\enddata
\end{deluxetable}

\begin{deluxetable}{llllllll}
\tabletypesize{\scriptsize}
\tablewidth{0pt}
\tablecaption{{\it Spitzer} Photometry for Candidate Members of 
Chamaeleon I\label{tab:cand}}
\tablehead{
\colhead{Name} &
\colhead{[3.6]} &
\colhead{[4.5]} &
\colhead{[5.8]} &
\colhead{[8.0]} &
\colhead{Date} &
\colhead{[24]} &
\colhead{Date} 
}
\startdata
2M J11020610$-$7718079 & 13.42$\pm$0.02 & 13.10$\pm$0.02 & 12.84$\pm$0.03 & 12.34$\pm$0.03 & 2006 Jul 10 & out & \nodata \\
2M J11025374$-$7722561\tablenotemark{a} & 13.93$\pm$0.02 & 13.63$\pm$0.02 & 13.43$\pm$0.03 & 13.07$\pm$0.05 & 2006 Jul 10 &  9.80$\pm$0.07 & 2005 Feb 27 \\
               Cha J11062788$-$7724543\tablenotemark{b,c}  & 15.57$\pm$0.05 & 14.45$\pm$0.05 & 13.37$\pm$0.07 & 12.41$\pm$0.07 & 2004 Jun 10 &    8.89$\pm$0.20 & 2005 Feb 27 \\
               Cha J11062854$-$7618039 & 14.21$\pm$0.02 & 13.89$\pm$0.02 & 13.62$\pm$0.03 & 13.15$\pm$0.05 & 2006 Jul 10 & 10.27$\pm$0.15 & 2004 Apr 11 \\
               Cha J11080397$-$7730382\tablenotemark{c} & 15.31$\pm$0.02 & 14.50$\pm$0.02 & 13.80$\pm$0.03 & 12.72$\pm$0.03 & 2004 Feb 19 &  7.82$\pm$0.03 & 2005 Feb 27 \\
                            & 15.28$\pm$0.03 & 14.48$\pm$0.02 & 14.03$\pm$0.08 & 12.80$\pm$0.07 & 2004 Jun 10 &                &            \\
2M J11085367$-$7521359\tablenotemark{d}  &  8.70$\pm$0.02 &  8.48$\pm$0.02 &  8.32$\pm$0.03 &  7.96$\pm$0.03 & 2004 Jul 21 &      4.98$\pm$0.03          &  2005 Feb 27          \\
2M J11100336$-$7633111\tablenotemark{c,e} & 12.74$\pm$0.02 & 11.98$\pm$0.02 & 11.51$\pm$0.04 & 10.94$\pm$0.08 & 2004 Jul 4 &  5.28$\pm$0.10 & 2004 Apr 11 \\
\enddata
\tablenotetext{a}{[LES2004] 424.}
\tablenotetext{b}{$7\arcsec$ from 2M~J11062942$-$7724586.}
\tablenotetext{c}{Possible class I source.}
\tablenotetext{d}{$2\arcmin$ from RX~J1108.8$-$7519A and B.} 
\tablenotetext{e}{OTS 32. $18\arcsec$ from Hn 11.}
\tablenotetext{f}{$8\arcsec$ from RX~J1129.2$-$7546.}
\end{deluxetable}

\begin{deluxetable}{lllllllll}
\tabletypesize{\scriptsize}
\tablewidth{0pt}
\tablecaption{{\it Spitzer} Photometry for Known Members of 
Chamaeleon I\label{tab:mem}}
\tablehead{
\colhead{2MASS\tablenotemark{a}} &
\colhead{Name} &
\colhead{[3.6]} &
\colhead{[4.5]} &
\colhead{[5.8]} &
\colhead{[8.0]} &
\colhead{Date} &
\colhead{[24]} &
\colhead{Date} 
}
\startdata
J10523694$-$7440287 &         \nodata & 10.19$\pm$0.02 & 10.11$\pm$0.02 & 10.08$\pm$0.03 & 10.14$\pm$0.04 & 2006 Jul 10 &  9.44$\pm$0.15 & 2007 Jul 16 \\
J10580597$-$7711501 &         \nodata & 11.70$\pm$0.02 & 11.45$\pm$0.02 & 11.26$\pm$0.03 & 10.75$\pm$0.04 & 2006 Jul 10 &            out &         \nodata \\
J10581677$-$7717170 &         T6 &            sat &            out &  6.41$\pm$0.03 &            out & 2006 Jul 10 &            out &         \nodata \\
J11011370$-$7722387 &         \nodata & 11.69$\pm$0.02 & 11.61$\pm$0.02 & 11.60$\pm$0.03 & 11.56$\pm$0.04 & 2006 Jul 10 &            \nodata & 2005 Feb 27 \\
J11011926$-$7732383 &         \nodata &            out & 10.87$\pm$0.02 &            out & 10.77$\pm$0.04 & 2004 Jul 4 &            \nodata & 2004 Jun 22 \\
                 &            & 11.00$\pm$0.02 & 10.86$\pm$0.02 & 10.81$\pm$0.03 & 10.76$\pm$0.04 & 2005 May 9 &                &            \\
                 &            & 11.00$\pm$0.02 &            out & 10.78$\pm$0.03 &            out & 2006 Jul 10 &                &            \\
J11013205$-$7718249 &    ESO H$\alpha$ 554 & 13.05$\pm$0.02 & 12.97$\pm$0.02 & 12.89$\pm$0.03 & 12.88$\pm$0.05 & 2006 Jul 10 &            out &         \nodata \\
\enddata
\tablecomments{Entries of ``$\cdots$", ``sat", ``out", and ``bad" 
indicate measurements that are absent because of non-detection, 
saturation, a position outside the field of view of the filter, and 
contamination by a cosmic ray or a ghost from a bright star, respectively.}
\tablenotetext{a}{2MASS Point Source Catalog.}
\end{deluxetable}

\begin{deluxetable}{lrrrrr}
\tabletypesize{\scriptsize}
\tablewidth{0pt}
\tablecaption{Spectral Slopes for Known Members of 
Chamaeleon I\label{tab:alpha}}
\tablehead{
\colhead{Name} &
\colhead{$\alpha$(2-8~\micron)} &
\colhead{$\alpha$(2-24~\micron)} &
\colhead{$\alpha$(3.6-8~\micron)} &
\colhead{$\alpha$(3.6-24~\micron)} &
\colhead{SED Class} 
}
\startdata
2M J10523694$-$7440287 & $-$2.60 & $-$2.51 & $-$2.81 & $-$2.58 & III \\
2M J10580597$-$7711501 & $-$1.76 &    \nodata & $-$1.78 &    \nodata &  II \\
2M J11011370$-$7722387 & $-$2.45 &    \nodata & $-$2.73 &    \nodata & III \\
2M J11011926$-$7732383 & $-$2.24 &    \nodata & $-$2.62 &    \nodata & III \\
2M J11013205$-$7718249 & $-$2.31 &    \nodata & $-$2.65 &    \nodata & III \\
2M J11021927$-$7536576 & $-$2.55 & $-$2.56 & $-$2.77 & $-$2.66 & III \\
2M J11022491$-$7733357 & $-$2.57 & $-$0.72 & $-$2.54 & $-$0.24 &  II \\
2M J11022610$-$7502407 & $-$2.51 & $-$2.50 & $-$2.67 & $-$2.57 & III \\
\enddata
\tablenotetext{a}{Spectral slopes are the observed values without a
correction for extinction.}
\end{deluxetable}

\begin{deluxetable}{lrrr}
\tabletypesize{\scriptsize}
\tablewidth{0pt}
\tablecaption{SED Classes in Chamaeleon~I\label{tab:class}}
\tablehead{
\colhead{Class} &
\colhead{North\tablenotemark{a}} &
\colhead{South\tablenotemark{b}} &
\colhead{Total}}
\startdata
I & 1 & 3 & 4 \\
flat & 2 & 8 & 10 \\
II & 43 & 51 & 94 \\
III & 44 & 51 & 95 \\
\enddata
\tablenotetext{a}{$\delta>-77\arcdeg$.}
\tablenotetext{b}{$\delta<-77\arcdeg$.}
\end{deluxetable}

\begin{deluxetable}{lcc}
\tabletypesize{\scriptsize}
\tablewidth{0pt}
\tablecaption{Disk Fractions\label{tab:diskfraction}}
\tablehead{
\colhead{Spectral Type} &
\colhead{Chamaeleon I} &
\colhead{IC 348}}
\startdata
$<$K6 & 14/19 = $0.74^{+0.08}_{-0.12}$	& 7/35 = $0.20^{+0.08}_{-0.05}$ \\
$\geq$K6 and $\leq$M3.5 &  42/66 = $0.64\pm0.06$ & 41/94 = $0.44\pm0.05$ \\
$>$M3.5 and $<$M6 & 36/83 = $0.43\pm0.05$  & 62/130 = $0.48\pm0.04$ \\ 
$\geq$M6 and $\leq$M8 & 10/24 = $0.42^{+0.1}_{-0.09}$ & 13/23 = $0.57\pm0.1$ \\
$>$M8 & 5/10 = $0.50^{+0.14}_{-0.15}$ & 5/12 = $0.42^{+0.14}_{-0.12}$ \\
\enddata
\tablecomments{The statistical errors in the disk fractions are computed in the
manner described by \citet{bur03}.}
\end{deluxetable}

\begin{deluxetable}{llllllllll}
\tabletypesize{\scriptsize}
\tablewidth{0pt}
\tablecaption{{\it Spitzer} Photometry for Probable Members of 
$\epsilon$~Cha\label{tab:eps}}
\tablehead{
\colhead{ID\tablenotemark{a}} &
\colhead{2MASS\tablenotemark{b}} &
\colhead{Name} &
\colhead{[3.6]} &
\colhead{[4.5]} &
\colhead{[5.8]} &
\colhead{[8.0]} &
\colhead{Date} &
\colhead{[24]} &
\colhead{Date} 
}
\startdata
13 & J11183572$-$7935548 &         \nodata &  9.31$\pm$0.02 &  9.15$\pm$0.02 &  9.06$\pm$0.03 &  8.50$\pm$0.04 & 2006 Jul 6 &            out &         \nodata \\
14 & J11225562$-$7924438 & RX J1123.2$-$7924 &  9.45$\pm$0.02 &  9.46$\pm$0.02 &  9.42$\pm$0.03 &  9.40$\pm$0.03 & 2004 May 1 &  9.53$\pm$0.10 & 2004 Aug 3 \\
15 & J11334926$-$7618399 &         \nodata & 10.92$\pm$0.02 & 10.84$\pm$0.02 & 10.82$\pm$0.03 & 10.79$\pm$0.03 & 2006 Jul 10 &  9.95$\pm$0.27 & 2007 Jun 6 \\
16 & J11404967$-$7459394 &         \nodata & 11.38$\pm$0.02 & 11.28$\pm$0.02 & 11.24$\pm$0.03 & 11.17$\pm$0.03 & 2006 Jul 10 &            \nodata & 2007 Jul 16 \\
17 & J11432669$-$7804454 &         \nodata & 10.08$\pm$0.02 &  9.87$\pm$0.02 &  9.67$\pm$0.03 &  9.26$\pm$0.03 & 2006 Jul 10 &            out &         \nodata \\
18 & J11493184$-$7851011 & RX J1149.8$-$7850 &  7.96$\pm$0.02 &  7.53$\pm$0.02 &  7.01$\pm$0.03 &  5.81$\pm$0.03 & 2004 Jun 10 &  1.75$\pm$0.04 & 2005 Mar 6 \\
19 & J11502829$-$7704380 & RX J1150.4$-$7704 &  8.86$\pm$0.02 &  8.86$\pm$0.02 &  8.85$\pm$0.03 &  8.82$\pm$0.04 & 2004 May 26 &  8.76$\pm$0.06 & 2005 Feb 27 \\
20 & J11582681$-$7754450 & RX J1158.5$-$7754B &  9.25$\pm$0.02 &  9.20$\pm$0.02 &  9.16$\pm$0.03 &  9.15$\pm$0.04 & 2004 Jun 10 &  8.61$\pm$0.06 & 2004 Feb 22 \\
\enddata
\tablenotetext{a}{Identifications assigned in this work, which conform to 
the ones used for previously known members of $\epsilon$~Cha 
\citep{fei03,luh04eta}.}
\tablenotetext{b}{2MASS Point Source Catalog.}
\end{deluxetable}

\begin{deluxetable}{lrrrrr}
\tabletypesize{\scriptsize}
\tablewidth{0pt}
\tablecaption{Spectral Slopes for Probable Members of 
$\epsilon$~Cha\label{tab:eps2}}
\tablehead{
\colhead{ID\tablenotemark{a}} &
\colhead{$\alpha$(2-8~\micron)} &
\colhead{$\alpha$(2-24~\micron)} &
\colhead{$\alpha$(3.6-8~\micron)} &
\colhead{$\alpha$(3.6-24~\micron)} &
\colhead{SED Class} 
}
\startdata
13 & $-$1.99 &    \nodata & $-$1.92 &    \nodata &  II \\
14 & $-$2.60 & $-$2.83 & $-$2.80 & $-$2.97 & III \\
15 & $-$2.54 & $-$2.43 & $-$2.71 & $-$2.47 & III \\
16 & $-$2.37 &    \nodata & $-$2.62 &    \nodata & III \\
17 & $-$1.87 &    \nodata & $-$1.92 &    \nodata &  II \\
18 & $-$0.86 & $-$0.28 & $-$0.37 &  0.08 &  flat \\
19 & $-$2.68 & $-$2.80 & $-$2.80 & $-$2.88 & III \\
20 & $-$2.59 & $-$2.57 & $-$2.74 & $-$2.62 & III \\
21 & $-$2.66 & $-$2.78 & $-$2.75 & $-$2.85 & III \\
\enddata
\tablenotetext{a}{Identifications from Table~\ref{tab:eps}.}
\end{deluxetable}

\clearpage

\end{document}